\pdfoutput=1
\documentclass[reprint,floatfix,superscriptaddress]{revtex4-2}
\usepackage{amsmath}
\usepackage{amssymb}
\usepackage{graphicx}
\usepackage{dcolumn}

\usepackage{xr}
\usepackage{cleveref}
\usepackage{bm}
\usepackage{xcolor}
\providecommand{\keywords}[1]
{
  \small	
  \textbf{\textit{Keywords---}} #1
}
\begin{document}

\date{\today}\title{Frequency-Dependent  Piezoresistive Effect \\ in  Top-down Fabricated Gold  Nanoresistors}

\newcommand{\BU}{Department of Mechanical Engineering, Division of Materials Science and Engineering, and the Photonics Center, Boston University, Boston, Massachusetts 02215, United States}

\author{C. Ti}
\affiliation{\BU}

\author{A. B. Ari}
\affiliation{\BU}

\author{M. C. Karakan}
\affiliation{\BU}

\author{C. Yanik}
\affiliation{SUNUM, Nanotechnology Research and Application Center, Sabanci University, Istanbul, 34956, Turkey}

\author{I. I. Kaya}
\affiliation{Faculty of Engineering and Natural Sciences, Sabanci University, Istanbul, 34956, Turkey}
\affiliation{SUNUM, Nanotechnology Research and Application Center, Sabanci University, Istanbul, 34956, Turkey}

\author{M. S. Hanay}
\affiliation{Department of Mechanical Engineering, Bilkent University, Ankara, 06800, Turkey}
\affiliation{National Nanotechnology Research Center (UNAM), Bilkent University, 06800, Ankara, Turkey}

\author{O. Svitelskiy}
\affiliation{Department of Physics, Gordon College, Wenham, Massachusetts 01984, United States}

\author{M. Gonz\'alez}
\affiliation{Aramco Services Company, Aramco Research Center--Houston, Houston, Texas, 77084, United States}

\author{H. Seren}
\affiliation{Aramco Services Company, Aramco Research Center--Houston, Houston, Texas, 77084, United States}

\author{K. L. Ekinci}
\email[Electronic mail: ]{ekinci@bu.edu}
\affiliation{\BU}

\date{\today}

\begin{abstract}

Piezoresistive strain gauges allow for electronic readout of  mechanical deformations with high fidelity. As piezoresistive strain gauges are aggressively being scaled down for applications in nanotechnology, it has  become  critical to investigate their physical attributes at different limits. Here, we describe an experimental approach for studying the piezoresistive gauge factor  of a gold thin-film nanoresistor as a function of frequency. The nanoresistor is fabricated lithographically near the anchor of a nanomechanical doubly-clamped beam resonator. As the resonator is driven to resonance in one of its normal modes, the nanoresistor is exposed to frequency-dependent strains  of {$\varepsilon \lesssim  10^{-5}$} in the $4-36~\rm MHz$  range. We calibrate the strain using optical interferometry and measure the resistance changes using a radio-frequency mix-down  technique. The piezoresistive gauge factor $\gamma$ of our lithographic gold nanoresistors  is $\gamma \approx 3.6$ at 4 MHz, in agreement with  comparable macroscopic thin metal film  resistors in previous works. However, our $\gamma$ values  increase monotonically with frequency and reach  $\gamma \approx 15$ at 36 MHz. We discuss  possible physics  that may give rise to this unexpected frequency dependence.

\end{abstract}

\keywords{Piezoresistive effect, Piezoresistive gauge factor, Gold nanowire, Gold  nanoresistor, NEMS}

\maketitle


The electrical resistance of a bar of metal or semiconductor is typically a function of the  mechanical strain on the bar, referred to as piezoresistivity or the piezoresistive effect \cite{fiorillo2018theory}.  By exploiting this change in resistance with strain,  a number of commonly-used  sensors have been developed for different technological applications and metrology. The fact that piezoresistive strain gauges are scalable in size has allowed for their integration into micro- and nano-electro-mechanical  systems (MEMS \cite{maluf2004introduction,villanueva2008crystalline,chui1998independent} and NEMS \cite{mile2010plane,li2007ultra,bargatin2007efficient, kouh2017nanomechanical}), paving the way for promising  technologies. The piezoresistive effect  is  quantified by the  gauge factor, $\gamma = \frac{1}{\varepsilon} \frac{\Delta R}{R}$, which typically relates the ``longitudinal strain" $\varepsilon$ to the fractional change in resistance, $\frac{\Delta R}{R}$ \cite{fiorillo2018theory,li2007ultra,bargatin2007efficient,tang2009metallic}. For  a simple resistor geometry such as a bar, one can write the resistance as $R = \frac{\rho L}{A}$ in terms of the resistivity $\rho$,   length $L$ and cross-sectional area $A$ of the resistor. This leads to the well-known expression for the gauge factor, $\gamma  = (1+2\nu) + \frac{1}{\varepsilon}\frac {\Delta \rho}{\rho}$, where $\nu$ is the Poisson's ratio \cite{fiorillo2018theory}. Thus,  two distinct mechanisms  determine the gauge factor. The first term, $(1+2\nu)$, typically less than 2, represents a purely geometric effect \cite{parker1963electrical}, i.e., an increase in length and a decrease in the cross-sectional area of the resistor. The second term, $\frac{1}{\varepsilon}\frac{\Delta \rho}{\rho}$, captures the changes in the intrinsic conduction of the material \cite{jen2003piezoresistance,neugebauer1962electrical}  arising from the  applied  strain. 

\begin{figure}
        \includegraphics[width=3. in]{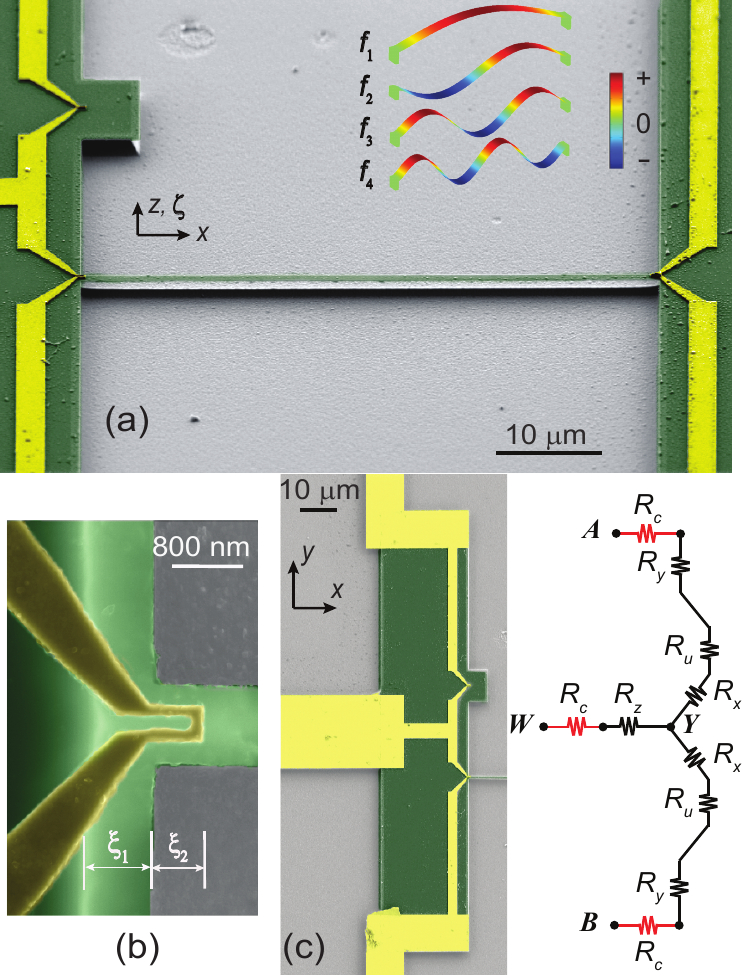}
    \caption{(a) SEM image of a silicon  nitride  doubly-clamped beam with linear dimensions of $l\times w \times t \approx \rm 50~\mu m \times 900~nm \times 100~nm$. The  gold nanoresistors fabricated on the anchors act as a strain gauge  (left) and an electrothermal actuator  (right). The eigen-modes of the structure are shown in the finite element simulations in the inset with the color bar corresponding to the out-of-plane displacement. (b) A close-up of the  strain gauge. This u-shaped thin film nanoresistor  has a thickness of $h=135$ nm and a width of 120 nm. The undercut region ($\xi_1 \approx 800$ nm) and the region on the bridge ($\xi_2 \approx 600$ nm) are both deformed when the beam vibrates. (c)  SEM image of the electrode (left) and the circuit model (right). The strain gauge is balanced with a nominally  identical  nanoresistor.  The  resistance of the strain gauge and the balancing resistor is $R_u$;  $R_x$, $R_y$, and $R_z$ are the  resistances of the lithographic  wires in different regions of the electrode; $R_{c}$ is the (average) electrical contact resistance from wirebonds.}
    \label{Figure1}
\end{figure}
 
In this manuscript, our focus is on  the piezoresistivity of  technologically-important thin metal-film resistors. The piezoresistive properties of  metal films have been investigated extensively  as a function of  sheet resistance (thickness)\cite{parker1963electrical}, strain \cite{verma1970strain,verma1972size}, and structure (i.e., grain size and separation)\cite{tellier1977grain}. In thicker films with low sheet resistances $R_s\lesssim 10^3~\Omega/\square$, the geometric effect dominates, resulting in $\gamma \lesssim 5$. The deviation of $\gamma$ from the purely geometric limit of $\gamma \approx 2$ has been attributed to the increase in the vibrational amplitude of the crystal atoms due to the applied strain; this results in the Gruneisen constant $G$  to enter the expression for $\gamma$ \cite{kuczynski1954effect,parker1963electrical,tellier1977grain} as $\gamma=(1+2\nu) + \left[ 1 + 2G(1-2\nu) \right ]$. In ultra-thin  films with large sheet resistances, $\gamma$  can easily exceed $10^3$ \cite{parker1963electrical,jen2003piezoresistance}, suggesting that  electron tunneling between grains and through cracks in the film become relevant. Most of the aforementioned measurements of $\gamma$ have been performed using static or relatively low-frequency ($\lesssim 1$ MHz) strains  \cite{li1994thin} --- even though strain gauges have been used at frequencies  higher than 100 MHz \cite{li2007ultra}.  

Looking at all the previous body of work on piezoresistivity of metal films, studies on  two important limits remain missing. The first is the piezoresistivity of a metal resistor with nanoscale cross-sectional dimensions. The few studies on nanowires are based on semiconducting nanowires \cite{he2008self,vietadao2015piezoresistive,neuzil2010electrically}. Second, the frequency dependence of the piezoresistive effect, at the nanoscale or otherwise, has not yet been addressed methodically, possibly due to measurement challenges. Here, we address these questions by measuring the gauge factor of a nanoscale strain gauge as a function of frequency at room temperature. We show that $\gamma$ of our nanoscale strain gauge at 4 MHz agrees with previous reports on macroscopic gold films at low frequency, suggesting that conduction in our nanoresistor is similar to that in macroscopic films.   Our $\gamma$ values, however, increase monotonically with frequency, reaching $\gamma \approx 15$ at 36 MHz. 

We perform our study of piezoresistivity of  nanoresistors using NEMS resonators such as the one shown in the scanning electron microscope (SEM) image in Fig. \ref{Figure1}(a). This is a tension-dominated silicon nitride doubly-clamped beam with linear dimensions of $l\times w \times t \approx \rm 50~\mu m \times 900~nm \times 100~nm$. On the two anchor regions of the doubly-clamped beam,  gold electrodes are patterned   using electron beam lithography, thermal film deposition, and lift off. The strain gauge is shown in  Fig. \ref{Figure1}(b): this is a 135-nm-thick lithographic u-shaped gold nanowire  and is fabricated over the anchor region of the suspended silicon nitride beam  (the brighter region in the SEM image in Fig. \ref{Figure1}(b)). The strain gauge is ``wired" into a bridge circuit  along with a nominally identical nanoresistor, as shown in Fig. \ref{Figure1}(c). The circuit diagram in Fig. \ref{Figure1}(c) represents the entire bridge circuit embedding the  strain gauge and the balancing resistor. Here, $R_{u}$ corresponds to  the resistances of the strain gauge and the balancing resistor; $R_x$, $R_{y}$,  $R_{z}$ are the lithographic  wires connecting the nanoresistors to three mm-scale wirebonding pads; the contact resistances  $R_{c}$ correspond to the wirebonds \cite{Supplementary2021Frequency-Dependent}. The resistance values for the circuit elements in Fig.~\ref{Figure1}(c) are calculated from the experimentally-measured resistivity $\rho$ of the gold film. To this end, we first make a four-wire measurement of the gold resistor represented by $R_y+R_u+R_x+R_z$ and find this resistance to be $14.51\pm 0.14 ~\Omega$. We then compute the same resistance from geometry (i.e., SEM images) in terms of an unknown $\rho$ using two methods: i) we integrate the infinitesimal resistance $dR = \rho \frac{d{{\ell}}}{  {h} W(\ell)}$ along the electron path using  the position-dependent width $W(\ell)$; ii) we ``count" the number of squares $N$ and determine the total resistance as $N \frac{\rho}{h}$. We find $\rho \approx 2.82 \times 10^{-8}~ \rm \Omega \cdot m$.  With $\rho$ determined, we calculate the resistance of each individual resistor from its geometry, as reported in Table \ref{Resistance}. There is typically a small mismatch between the two arms of the bridge of about $2-5\%$, which contributes to the errors. Subsequent two-wire measurements provide the contact resistances of the wirebonds to be $R_{c}= 1.18 \pm 0.07~ \Omega$.  On the second anchor of the NEMS beam (right anchor in Fig.~\ref{Figure1}(a)), an identical nanoresistor is fabricated for electrothermal actuation of nanomechanical oscillations \cite{bargatin2007efficient}. 

Our overall approach is as follows. We drive the resonator in several of its  eigen-modes shown in the upper inset of Fig.~\ref{Figure1}(a) using the electrothermal actuator. The oscillation amplitude of the resonator is carefully calibrated as a function of the drive voltage applied to the electrothemal actuator in a heterodyne optical interferometer with a displacement noise floor of $\sim 20~ \rm fm/Hz^{1/2}$ at a sample power of $100~\rm \mu W$.  In separate electrical measurements, the piezoresistance  is measured during calibrated eigen-mode oscillations. From the oscillation amplitude,  the longitudinal strain is calculated numerically and  the gauge factor is extracted as a function of (eigen-mode) frequency. We have measured three devices from the same batch with identical strain gauges and embedding circuits (Figs. \ref{Figure1}(b) and (c)), a 60-$\rm \mu m $-long device, a 50-$\rm \mu m $-long device, and a 30-$\rm \mu m $-long device, with all the relevant parameters listed in Table \ref{tab_devices}. Further experimental details are provided in the SI \cite{Supplementary2021Frequency-Dependent}.

\begin{figure}
  
    \includegraphics[width=3.375in]{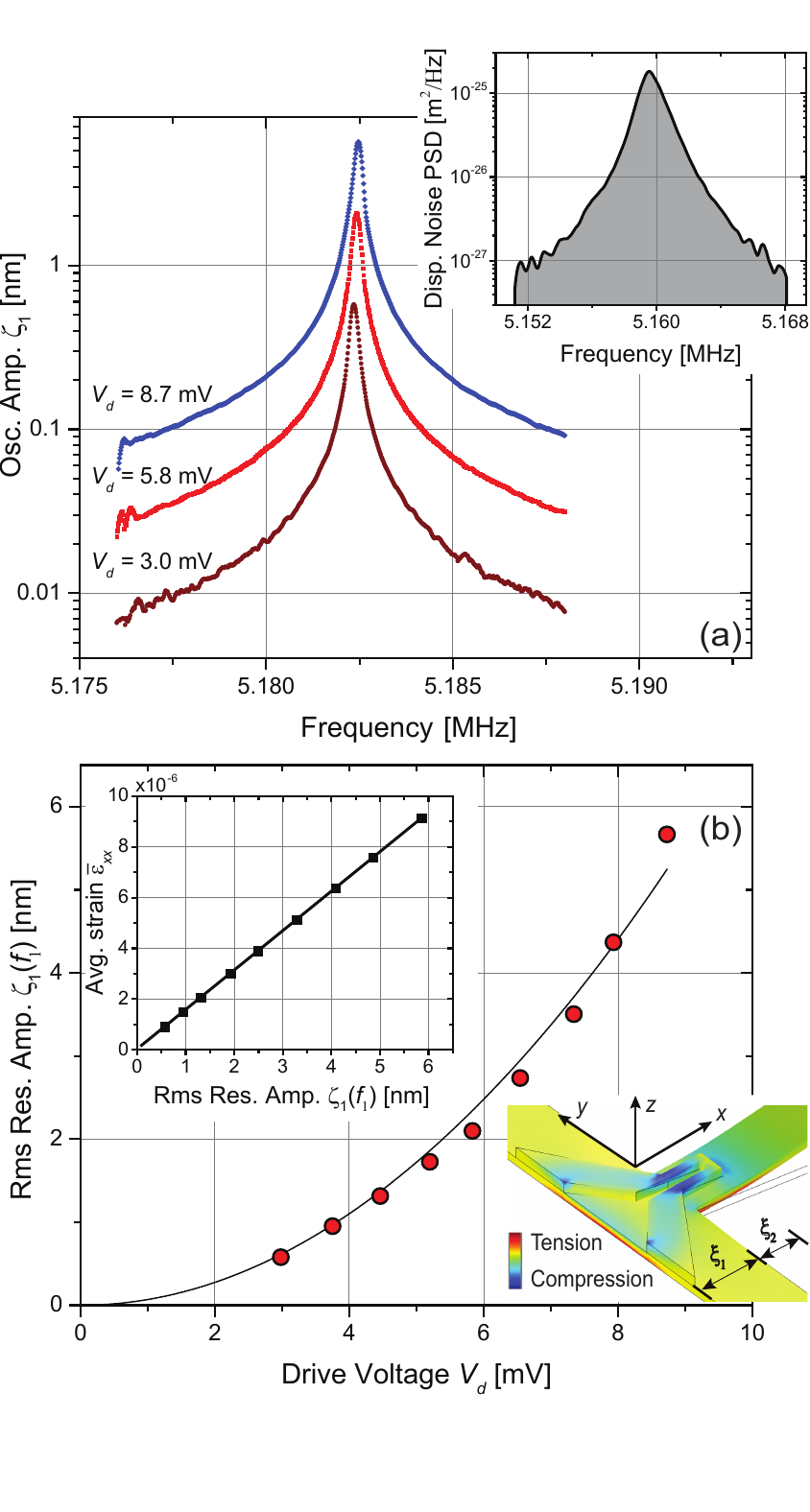}
    \caption{ (a) Rms oscillation amplitude  of a 50-$\rm \mu m$-long beam at different drives around its  fundamental mode resonance frequency measured at its center ($x = l/2$).  (Inset) power spectral density (PSD) of the Brownian fluctuations of the same mode. (b)  Rms resonance amplitude $\zeta_1(f_1)$ of the fundamental mode plotted as a  function of the drive voltage $V_d$. The continuous line is a fit to $\zeta_1 = A_1 {V_d} ^2$. The lower inset is from a finite element model (FEM) showing the relevant strain field ${\varepsilon}_{xx} (\mathbf{r})$. The upper inset is the average strain $\bar \varepsilon_{xx}$ as a function of  resonance amplitude $\zeta_1(f_1)$ determined from FEM  such as the one shown in the lower inset. The strain is linear with amplitude, $\bar \varepsilon_{xx} = \chi_1 \zeta_1(f_1) $, with a slope of $1.6 \times 10^{-6} ~{\rm nm}^{-1}$.}
     \label{Figure2}
\end{figure}
\begingroup
\setlength{\tabcolsep}{10pt} 
\renewcommand{\arraystretch}{1.2} 
\begin{table}
\caption{Resistance values for each lithographic resistor in the device  calculated from the resistivity $\rho$  and the geometry. The typical error in these values is 5$\%$.} \label{Resistance}

\begin{tabular}{ccccc}
 $R_{x}$ &  $R_{y}$  &  $R_{u}$  &   $R_{z}$  &  $R_{c}$ \\
 
 \hline
 $2.16~\Omega$  &  $7.37~ \Omega$   & $3.54 ~\Omega $  &  $1.43~\Omega $   &  $1.18 ~\Omega$   \\
\end{tabular}
\end{table}
\endgroup


\begingroup
\setlength{\tabcolsep}{10pt} 
\renewcommand{\arraystretch}{1} 
\begin{table}
\caption{ Experimentally-obtained mechanical properties of the measured devices. } \label{tab_devices}

\begin{tabular}{cccccccccc}
$l \times w \times t$ & $n$ & $f_n$  & $k_n$ & $\chi_n$ \\
($\mu \rm m^3$)  &   & (MHz) &   (N/m)  &($\times 10^{-6}\rm/nm$)  \\
\hline
 $60 \times 0.90 \times 0.1$ &  1 & 4.3  & 6.3  & 1.3 \\
                               &  2 & 8.9  & 23.2   & 2.7\\
                              &  3 & 12.9  & 55.8   & 3.9 \\
                               &  4 & 17.3  & 93.8  & 5.2 \\
$50 \times 0.90 \times 0.1$ &  1 & 5.2  & 7.4   &  1.6 \\
                               &  2 & 10.4  & 29.3  &  3.1 \\
                              &  3 & 15.6  & 69.0   &  4.5 \\
                              &  4 & 20.8  & 125.0    &  6.1 \\ 
$30 \times 0.90 \times 0.1$ &  1 & 8.8  & 11.8   &  2.7 \\
                               & 2 & 17.7  & 45  &  5.3 \\
                              &  3 & 26.6  & 98   &  8.1 \\
                              &  4 & 35.6  & 192    &  10.6
\end{tabular}
\end{table}
\endgroup


We illustrate the optical calibration of the strain for the fundamental mode of the 50-$\rm \mu m $ resonator. The resonance curves for the  mode are shown in Fig. \ref{Figure2}(a). Here,  the electrothermal actuator excites the nanomechancial resonance with a harmonic force at different rms amplitudes, with the frequency of the drive force  swept around the fundamental mode resonance frequency $f_1$ (frequency of the electrical drive swept around $f_1/2$). The rms oscillation amplitude $\zeta_1$ is measured optically at the anti-node (i.e., the center). From this measurement, we obtain the mode resonance frequency  and  quality  factor  as $f_1 \approx 5.1825~\rm MHz$ and $Q_1\approx 2.9 \times 10^4$, respectively. The inset shows the power spectral density (PSD) of the thermal fluctuations of the same mode of a  nominally identical beam, with the integral of the PSD providing the spring constant $k_1\approx \rm 7.4 ~N/m $ from the equipartition of energy \cite{ari2020nanomechanical}. The mechanical parameters in Table \ref{tab_devices} are obtained from similar measurements on other modes, with all the data presented in the SI \cite{Supplementary2021Frequency-Dependent}. Since the measurements are performed in a vacuum chamber, the  quality factors are dependent on the residual pressure in the chamber and are typically high ($5 \times 10^2 \lesssim Q \lesssim 2 \times 10^4$).  The effect of the  $Q$ factor on the measurements are properly removed as discussed below. 
\begin{figure}
    \centering
    \includegraphics[width=3.375in]{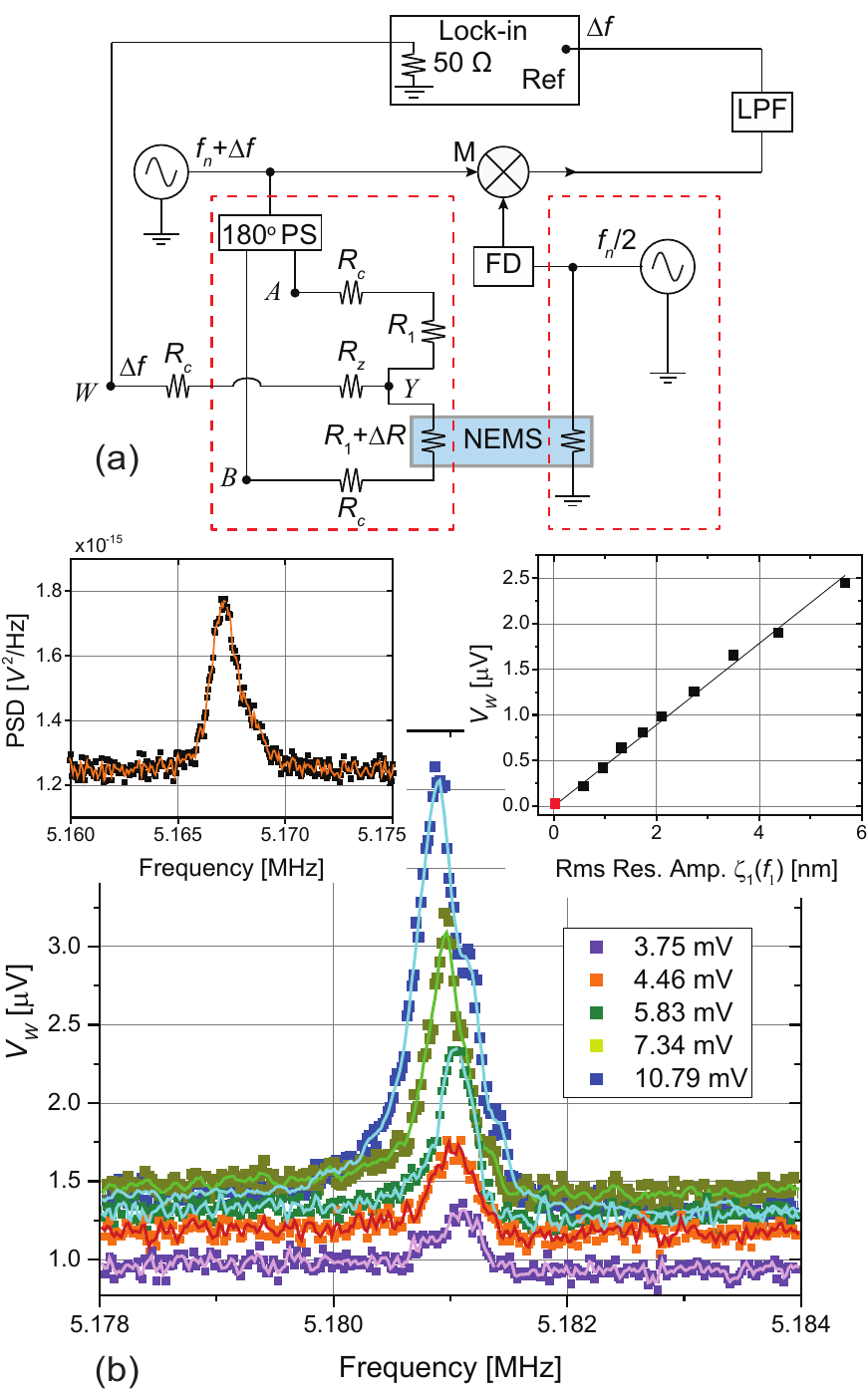}
    \caption{ (a) Simplified schematic diagram of the mix-down measurement of piezoresistance. 180$^{\rm o}$ PS: 180$^{\rm o}$ power splitter; LPF: low pass filter; FD: frequency doubler. (b) Piezoresistance  signal $V_{W}$ at different drives as a function of frequency around the fundamental resonance of the 50-$\rm \mu m$-long resonator. The  bias voltage, $V_b = 60 ~\rm mV$, is kept constant for all curves. Left inset shows the PSD of the  Brownian  fluctuations  of the mode coupling to the  piezoresistance signal. Right inset is the measured $V_{W}$ as a function of resonance amplitude of the resonator; all quantities are rms. The red data point is the signal from the Brownian motion. The line is a linear fit through the origin. The error bars in the right inset are smaller than the symbols.}
    \label{Figure3}
\end{figure}

Fig. \ref{Figure2}(b) shows the rms resonance amplitude $\zeta_1(f_1)$ as a function of the drive voltage $V_d$ for the fundamental mode of the  $ 50$-$\rm \mu m$-long resonator. These data are essentially the peak values of the resonance curves, such as those shown in Fig. \ref{Figure2}(a). The solid line in Fig. \ref{Figure2}(b)  is a fit of the form $\zeta_1 = A_1 {V_d}^2$. The parabolic dependence on voltage arises from the physics of the electrothermal actuator \cite{bargatin2007efficient}. The upper inset of Fig. \ref{Figure2}(b) shows the average longitudinal strain ${\bar \varepsilon}_{xx}$ on the nanoresistor as a function of the resonance amplitude $\zeta_1(f_1)$. To find the  strain in the  nanoresistor due to the bending of  the  silicon nitride structure,  we have resorted to the finite element method (FEM).  {Briefly, we solve for the eigen-frequencies of the resonator (including the undercut regions) using  boundary mode analysis. Since the resonator is under tension, we apply a  tensile load to the silicon nitride layer to match the simulated and  experimental eigen-frequencies. We then impose an rms displacement amplitude for the beam at its anti-node  and calculate the corresponding strain field.}  Lower inset of Fig. \ref{Figure2}(b)  shows the rms longitudinal strain field $\varepsilon_{xx}(\mathbf{r})$  as a function of position $\mathbf{r}$ over the suspended base region  for an (rms)  resonance amplitude of  $\zeta_1(f_1) \approx 7 \rm ~nm$ (at the center) for the  50-$\rm \mu m$ beam in its fundamental mode. To calculate the average value of $\varepsilon_{xx}(\mathbf{r})$, we first average over the cross-sectional area parallel to  the $yz$ plane, ${S_{yz}(x)}$, of the nanoresistor, finding  $ {\varepsilon}_{xx}(x) = \frac{1}{S_{yz}(x)}{\iint\limits_{S_{yz}(x)} \! {\varepsilon}_{xx} (\mathbf{r}) dydz}$. We ignore the contribution from the small nanoresistor region that is parallel to the $y$ axis. Next, we average $ {\varepsilon}_{xx}(x)$ along the length of the nanoresistor (i.e., $x$ axis). With the origin  at the position where the beam structure starts,  $\bar \varepsilon_{xx} = {1 \over {\xi_1  + \xi_2 }}\int\limits_{ - \xi_1 }^{\xi_2}  {{\varepsilon _{xx}}(x)dx}$. The linear dimensions $\xi_1$ and $\xi_2$ are shown in Fig. \ref{Figure1}(b) and the lower inset of \ref{Figure2}(b).  As a result, we find that, for all modes,   $\bar \varepsilon_{xx}$   depends linearly  on the resonance amplitude of the resonator as $\bar \varepsilon_{xx} = \chi_n \zeta_n(f_n)$ where $\chi_n$ is a constant. The results for the fundamental mode of the 50-$\rm \mu m$ beam are shown in the upper inset of Fig.~\ref{Figure2}(b), and all the values of $\chi_n$ are listed in Table \ref{tab_devices}. 

Now, we turn to the measurement of the piezoresistance signal  during   driven fundamental eigen-mode oscillations. To reduce  parasitic effects, we employ a mix-down measurement \cite{bargatin2005sensitive}  in the balanced circuit \cite{ti2020optimization}  shown in Fig.~\ref{Figure3}(a).  Briefly, the resonator is driven at its resonance at $f_n$ by applying a voltage at $\frac{f_n}{2}$ to the electrothermal actuator \cite{bargatin2007efficient}, which generates temperature oscillations and hence a thermoelastic force at $f_n$. The  mechanical   strain  in the strain gauge causes a time-varying piezoresistance   $ \sqrt{2} \Delta R \cos{(2\pi f_n t)}$. The mix-down and background reduction are accomplished by applying two out-of-phase  bias  voltages of  $\pm \sqrt{2} V_{b} \cos {(2\pi f_n t + 2\pi\Delta f t)}$  to the two arms of the bridge (ports $A$ and $B$ in Fig.~\ref{Figure3}(a) and Fig.~\ref{Figure1}(c)). Assuming  negligible imbalance in the bridge ($R_1 = R_2$) and $\Delta R \ll R_1$, we find the down-converted  signal at the input of the measurement electronics (at point $W$) as   $ \sqrt{2} V_{W} \cos{(2\pi\Delta f t)}$ where
\begin{equation}\label{eq 1}
    V_{W} = \frac{ V_b R_{t} \Delta R}{\sqrt{2} \left ( {R_1 + R_{c}+ 50~\Omega}\right)^2}.
\end{equation}
Here, $R_{t} = (\frac{2}{R_1+R_{c}+50} +  \frac{1}{R_{z} + R_{c} + 50})^{-1}$ with $R_1=R_{y}+R_{u}+R_{x}$. In our experiments, this signal in Eq.~(\ref{eq 1}) is detected using a lock-in amplifier referenced to ${\Delta f}= 1.5$ MHz; $V_b$ is kept constant. The  (rms) value of the piezoresistance $\Delta R$ is then found from the measured $ V_{W}$. From separate reflection measurements, we conclude that there is very little attenuation in the bias current $V_b/(R_1+R_{c}+50~\Omega)$ and hence the detected signal. The analysis of the detection circuit  and complementary measurements (e.g., reflection) are available in the SI \cite{Supplementary2021Frequency-Dependent}. 

\begin{figure*}
    \includegraphics[width=6.75in]{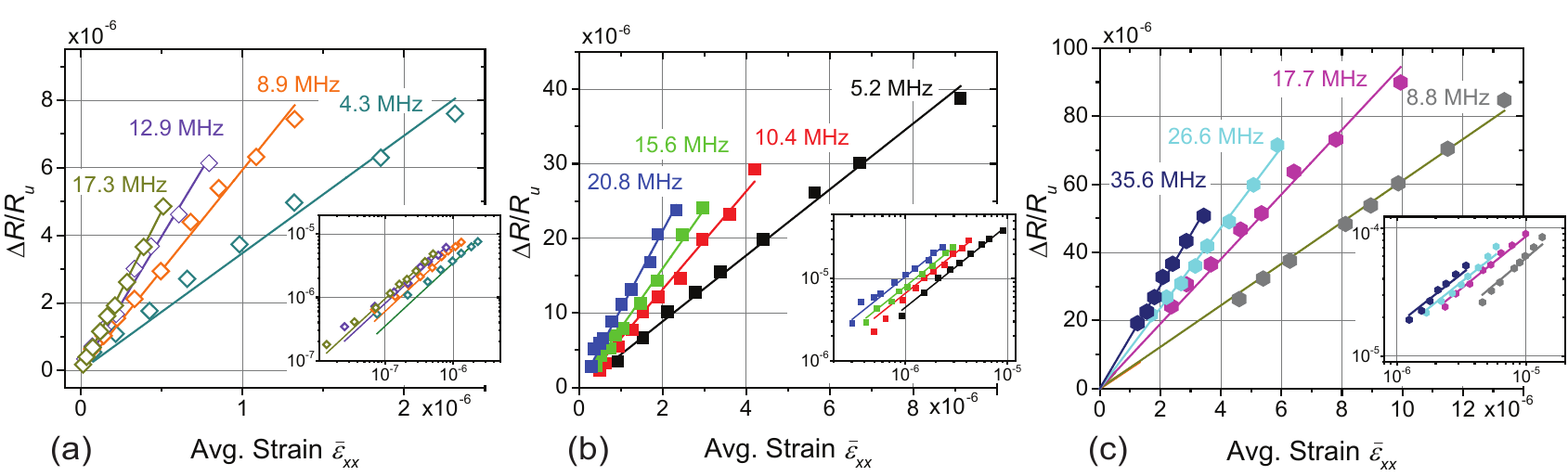}
    \caption{ $\frac{\Delta R}{R_{u}}$  as a function of  strain $\bar \varepsilon_{xx}$ for all the modes of (a) the 60-$\mu \rm m$, (b)  50-$\mu \rm m$,  and (c)   30-$\mu \rm m$ resonators. The insets in (a), (b),  and (c) show the same data in  double-logarithmic plots. The slopes of the linear fits provide $\gamma$. }
    \label{Figure5}
\end{figure*}

\begin{figure}
  
    \includegraphics[width=3.375in]{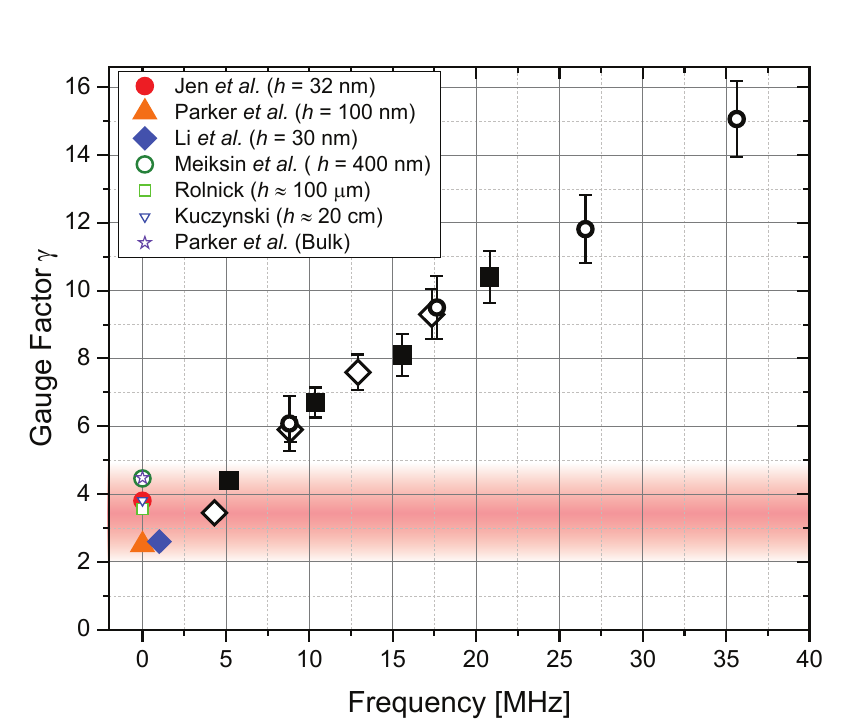}
    \caption{ Extracted  $\gamma$ as a function of (mode) frequency. The error bars are the rms errors in $\gamma$ \cite{Supplementary2021Frequency-Dependent}. The error bars at 4.3 MHz and 5.2 MHz are smaller than the symbols. For comparison, we also display previously-published $\gamma$ data obtained in quasi-static measurements. The filled  data points  are  for gold films with thicknesses and resistivities comparable to those of our films \cite{li1994thin,jen2003piezoresistance,parker1963electrical}; the open data points are obtained in  thick gold films or bulk gold. \cite{meiksin1967theoretical,rolnick1930tension,kuczynski1954effect,parker1963electrical}.}
     \label{Figure5_1}
\end{figure}

Fig.~\ref{Figure3}(b) shows the measured rms voltages $V_W$ on the strain gauge of the 50-$\rm \mu m $-long beam at different drives as  the drive frequency is swept through the  fundamental resonance. Compared with the optically detected resonance curves of Fig.~\ref{Figure2}(a), one notices that $f_1$ and $Q_1$ are slightly different. The left inset shows an electrical measurement of the  power spectral density (PSD) of the  thermal fluctuations of the resonator  on the strain gauge \cite{Supplementary2021Frequency-Dependent}.  The right inset shows the rms voltage due to piezoresistance as a function of the nanomechanical resonance amplitude. This voltage is determined by subtracting the baseline value from the peak value at $f_1$ in Fig. \ref{Figure3}(b). The resonance amplitude ($x$ axis) is determined from the optical calibration in Fig. \ref{Figure2} above after accounting for the  different $Q$ values in optical  and electrical measurements, e.g., due to  changes in the chamber pressure or resonator surface conditions. Since the amplitude $\zeta_1(f_1)$ is given by $\zeta_1(f_1) =  \frac{F_1 Q_1}{k_1}$ with $k_1$  a constant and $F_1$ only dependent on the applied external voltage, we scale $\zeta_1(f_1)$ by the ratio of the $Q$ factors under identical drive voltages.   We thus  obtain the data in the upper right inset of Fig. \ref{Figure3}(b). We add the thermal noise result (the red data point in the right inset of Fig. \ref{Figure3}(b))  to the $V_{W}$ \textit{vs.} amplitude plot with the understanding that both the voltage and the amplitude are rms quantities obtained from integrals of the PSD. With all the resistances and voltages in the circuit known, it is straightforward to compute the $\Delta R$ values. 

Finally, we extract the gauge factor by combining  all the measurements. For each mode, we convert the applied drive voltage into resonance amplitude using the optical calibration (Fig. \ref{Figure2}(b) main) and the amplitude into strain using the numerical simulations(Fig. \ref{Figure2}(b) inset). Figs. \ref{Figure5}(a), (b), and (c) show  $\frac{\Delta R}{R_{u}}$ for the strain gauge as a function of strain for the first four modes of the 60-$\rm \mu m $-long,   50-$\rm \mu m $-long, and 30-$\rm \mu m $-long  resonators, respectively. The insets show double-logarithmic plots of the same  data.   For all modes of the three resonators,  $\frac{\Delta R}{R_{u}}$ increases linearly with $\bar \varepsilon_{xx}$ with the slope being the gauge factor $\gamma$: $\frac{\Delta R}{R_{u}} =\gamma \bar \varepsilon_{xx}$.  We display  $\gamma$ for the 60-$\mu \rm m$,  50-$\mu \rm m$, and  30-$\rm \mu m $ resonators as a function of (mode) frequency in Fig. \ref{Figure5_1}. The gauge factor increases monotonically from 3.6 to 15 in the frequency range  $4.3 - 36~\rm MHz$.  The error bars in Fig. \ref{Figure5_1} represent rms errors  based on standard error analysis discussed in SI \cite{Supplementary2021Frequency-Dependent}. Also in Fig. \ref{Figure5_1}, we show  previously-published quasi-static $\gamma$ values for gold films of different thicknesses and for gold wires. The three filled data points around zero frequency show gold films \cite{li1994thin,jen2003piezoresistance,parker1963electrical} with  thicknesses and resistivitities (sheet resistances) comparable to those of our films. We also include four data points on very thick films and bulk gold (wires), shown by the open symbols  \cite{meiksin1967theoretical,rolnick1930tension,kuczynski1954effect,parker1963electrical}.  

Our lowest frequency $\gamma$ value is within the expected range and  close to those reported in the literature; this gives us confidence that our resonance-based measurements are  accurate. The   frequency dependent increase of our $\gamma$, however, is unexpected and cannot be traced to trivial  sources, e.g., heating or attenuation, which would cause an  effect in the opposite direction (i.e., a decrease in $\gamma$ with frequency). We therefore look for  possible  fundamental  mechanisms. The resistivity of our gold films, $\rho=2.81 \times 10^{-8} ~\rm \Omega \cdot m$,  is  close to that of very thick gold films ($\rho_B=2.44 \times 10^{-8} ~\rm \Omega  \cdot m$) \cite{jen2003piezoresistance,chopra1963electrical} and  bulk gold ($\rho_B=2.44 \times 10^{-8} ~\rm \Omega  \cdot m$) \cite{graz2009extended,goodman2002current,christie1994gold}. The slightly larger resistivity of our films compared to $\rho_B$ is possibly due to increased surface scattering. Regardless,  the electron relaxation time $\tau$ in our gold film at room temperature should be close to the bulk value of $\tau \sim 30 \times 10^{-15}$ s \cite{gall2016electron}. It seems unlikely that an electronic process is the source of the observed effect at the frequencies $f_n$ of our experiments, given that $f_n \tau \approx 0$.   We therefore speculate that mechanical effects  give rise to the observed increase. In particular,  it is possible that the resonant mechanical motion of the beam  couples to a mechanical mode of the nanoresistor or the grains within the nanoresistor. The average grain size in these films is 40 nm, and  gold nanorods and nanoparticles of similar dimensions have been shown to have acoustic resonances around 10-100 GHz \cite{zijlstra2008acoustic,pelton2009damping,ruijgrok2012damping}. Since the gold grains here are in a solid matrix and coupled to other grains mechanically, there could possibly exist lower-frequency mechanical modes within the thin film.  Hence, the mechanical energy of the beam may be coupling to these modes and actuating oscillatory  strains within the film larger than the strains predicted by FEM. These strains, in turn, may be increasing the grain to grain resistances, giving rise to the observed frequency dependence.  This is somewhat similar to the tunneling effects that have been discussed in the piezoresistivity of ultrathin films in which  grain to grain transport dominate the piezoresistance \cite{parker1963electrical}.  

In summary, we have described a method to measure the piezoresistive effect as a function of frequency. More experimental and theoretical studies  are needed to pinpoint the source of the observations here. In particular, increasing the frequency range may provide valuable insights. Also, repeating the  experiments on strain gauges with different linear dimensions and thicknesses and made up of different metals may help answer some of the questions. Regardless, the effect can be harnessed to develop efficient high-frequency NEMS devices.

\begin{center}
\textbf{Supporting Information}
\end{center}

Description of the measurement setup and device fabrication process; description of the optical  measurements, the procedure for the calibration of strains and  spring constants, optical data for all the modes of all devices; details of the electrical measurements; resistivity and RF measurements on the gold film electrodes; analysis of the electrical detection circuit; electrical data for all modes of all devices;  error analysis. 

\begin{acknowledgments}
We acknowledge support from Aramco Services Company (A-0208-2019) and the US NSF (CBET 1604075, CMMI 1934271, CMMI 2001403, DMR 1709282, and CMMI 1661700).
\end{acknowledgments}

\bibliography{Reference.bib}

  
    

\end{document}


\title{Supplementary Material for \\ ``Frequency-Dependent  Piezoresistive Effect \\ in  Top-down Fabricated Gold  Nanoresistors"}

\newcommand{\BU}{Department of Mechanical Engineering, Division of Materials Science and Engineering, and the Photonics Center, Boston University, Boston, Massachusetts 02215, USA}

\author{C. Ti}
\affiliation{\BU}

\author{A. Ari}
\affiliation{\BU}

\author{M. C. Karakan}
\affiliation{\BU}

\author{C. Yanik}
\affiliation{SUNUM, Nanotechnology Research and Application Center, Sabanci University, Istanbul, 34956, Turkey}

\author{I. I. Kaya}
\affiliation{Faculty of Engineering and Natural Sciences, Sabanci University, Istanbul, 34956, Turkey}
\affiliation{SUNUM, Nanotechnology Research and Application Center, Sabanci University, Istanbul, 34956, Turkey}

\author{M. S. Hanay}
\affiliation{Department of Mechanical Engineering, Bilkent University, Ankara, 06800, Turkey}
\affiliation{National Nanotechnology Research Center (UNAM), Bilkent University, 06800, Ankara, Turkey}

\author{O. Svitelskiy}
\affiliation{Department of Physics, Gordon College, Wenham, Massachusetts 01984, USA}

\author{M. Gonz\'alez}
\affiliation{Aramco Americas, Aramco Research Center--Houston, Houston, Texas, 77084, United States}

\author{H. Seren}
\affiliation{Aramco Americas, Aramco Research Center--Houston, Houston, Texas, 77084, United States}

\author{K. L. Ekinci}
\email[Electronic mail: ]{ekinci@bu.edu}
\affiliation{\BU}

\date{\today}

\maketitle

\tableofcontents

\newpage

\widetext

\setcounter{equation}{0}
\setcounter{figure}{0}
\setcounter{table}{0}
\setcounter{page}{1}
\makeatletter
\renewcommand{\theequation}{S\arabic{equation}}
\renewcommand{\thefigure}{S\arabic{figure}}
\renewcommand{\bibnumfmt}[1]{[S#1]}
\renewcommand{\citenumfont}[1]{S#1}

\section{Measurement Setup and Nanomechanical Devices}

\subsection{Measurement Chamber} \label{chamber pressure}

We use a vacuum chamber with optical and electrical access. The NEMS chip is glued to a chip carrier and wire bonds are attached to the chip. The chip carrier is placed in a chip socket that is soldered to a printed circuit board (PCB). The PCB is secured to the vacuum chamber with solder connections. The entire vacuum chamber is placed on a motorized $XYZ$ stage on an optical table. The stage is moved to keep the laser focused to the measurement location (i.e.,  the anti-node positions of the beam) for the optical measurements. The  pressure inside the chamber  is $p\gtrsim  10^{-5}$ Torr.  The measurements of the 30-$\rm \mu$m-long beam and the 50-$\rm \mu$m-long beam are performed in a vacuum close to the base pressure, whereas the measurement of the 60-$\rm \mu$m-long beam is completed at higher pressures in the several Torr range. The pressure  value in the vacuum chamber  strongly effects the $Q$ factor of the resonator \cite{kara2017generalized} but does not change the resonance frequency appreciably. We  remove the effects of $Q$ variations from our experimental measurements --- as described in  Section  \ref{freq spring Q}.

\subsection{Optical Interferometer}

\begin{figure}
    \centering
    \includegraphics[width=6.75in]{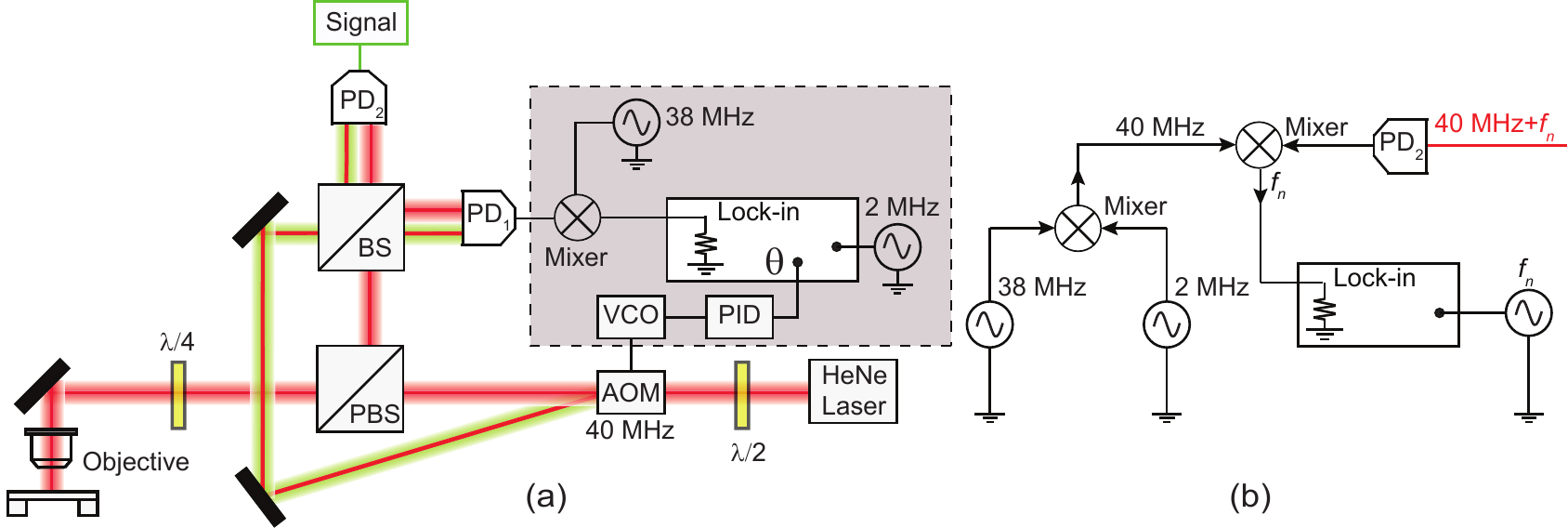}
    \caption{(a) Optical measurement setup. AOM: acousto-optic modulator, $\lambda/2$: half wave plate, $\lambda/4$: quarter wave plate, PBS: polarizing beam splitter, BS: beam splitter, PD: photodetector, PID: proportional-integral-derivative controller, VCO: voltage controlled oscillator. Signal on $\rm PD_1$ with components in the dashed box are used for feedback stabilization; $\rm PD_2$ is connected to a spectrum analyzer for  thermal noise measurements and to a lock-in amplifier  via a mixer for driven measurements. (b) Circuit diagram for driven measurements at resonance frequency $f_n$ of the resonator. }   
    \label{Figure Optsetup}
\end{figure}

We use a heterodyne Michelson interferometer \cite{lawall2000michelson} with feedback stabilization to detect the nanomechanical motion of the NEMS resonators  in both noise and driven measurements. Here, the optical signal is shifted up in frequency by $\rm 40~MHz$ using an acousto-optic modulator (AOM) to avoid  the low-frequency laser noise. The schematic diagram of the optical measurement setup with feedback stabilization is shown in Fig. \ref{Figure Optsetup}(a). A helium-neon laser with a wavelength of 632 nm is used in the frequency stabilization mode. The AOM diffracts the beam into two and modulates the diffracted arm. The interference signal is detected by two photodetectors, $\rm PD_1$  and $\rm PD_2$. Here, $\rm PD_1$ (Thorlabs PDA8A) is used to stabilize the feedback loop (dashed box) and $\rm PD_2$  (New Focus 1801)  is used to measure the nanomechanical motion of the NEMS resonator. In the experiments,  typical optical powers incident on a NEMS resonator and the photodiodes are 100 $\rm \mu W$ and 160 $\rm \mu W$, respectively. The   displacement sensitivity of the heterodyne interferometer is estimated to be $\sim \rm 20~ fm/\sqrt{Hz}$.  

The feedback stabilization (dashed box in Fig. \ref{Figure Optsetup}(a)) is for compensating for the low frequency phase noise due to any disturbances in the optical paths. The signal  from $\rm PD_1$ is first mixed down to 2 MHz and then demodulated by a   lock-in amplifier (SR865, Stanford Research Systems). The demodulated phase error signal is  fed to a proportional-integral-derivative (PID) controller, and the output of the PID drives a voltage controlled oscillator (VCO) which determines the drive frequency of the AOM. 

The displacement amplitude (oscillation amplitude) measured in the interferometer is calibrated against the wavelength of the HeNe laser as follows.  The calibration is based on the intensity signal detected on the two photodetectors, which is a result of the interference of the object beam (red in Fig. \ref{Figure Optsetup}(a)) with the reference beam (green in Fig. \ref{Figure Optsetup}(a)).  The intensity signal at each photodetector is mathematically described as $I_p = ({I_o} + {I_r}) \Big( 1 +  \frac{2\sqrt{I_o I_r}}{{I_o} + {I_r}} \cos{\big[2 \pi f_c t + k(z_r - z_o) + 2 k \delta}\big] \Big)$. Here, $I_o$ and $I_r$ are the intensities on the photodetectors resulting from the object and reference beams, respectively;  $f_c = 40$ MHz is the  modulation frequency;  $k = \frac{2 \pi}{\lambda}$ with $\lambda$ being the wavelength of the laser; $z_o$ and $z_r$ represent the optical paths traveled by the object and reference beams, respectively; $\delta$ is the vibration amplitude of the resonator. Since  $\delta$ is small, $I_p$ can be expanded using  trigonometric identities as 
\begin{equation}
   I_p = ({I_o} + {I_r}) \bigg[ 1 +  \frac{2\sqrt{I_o I_r}}{{I_o} + {I_r}} \Big(\cos{\big[2 \pi f_c t + k(z_r - z_o) }\big] - 2 k \delta \sin{\big[2 \pi f_c t + k(z_r - z_o)\big]}\Big)\bigg].
   \label{Intensity}
\end{equation} 
When the resonator is excited in one of its eigen-modes, $\delta (t) = \sqrt{2}\zeta_n \sin{2 \pi f_n t}$ with $\zeta_n$ and $f_n$ being the rms oscillation amplitude and the resonance frequency in the $n^{\rm th}$ mode. Using a Fourier-Bessel expansion, we obtain the intensity $I$ as \cite{wagner1990optical}
\begin{equation}
\begin{split}
      I_p =  & ({I_o} + {I_r}) \bigg[ 1 +  \frac{2\sqrt{I_o I_r}}{{I_o} + {I_r}} \Big(\cos{\big[2 \pi f_c t + k(z_r - z_o) \big]} + \\
      & k \zeta_n \cos{\big[2 \pi (f_c + f_n) t + k(z_r - z_o)\big] } - k \zeta_n \cos{\big[2 \pi (f_c - f_n) t + k(z_r - z_o)\big] }\Big)\bigg].
   \end{split}
   \label{disp cal}
\end{equation}
From Eq. \ref{disp cal}, the ratio $\mu$ of the sideband amplitudes at $f_c + f_n$ or $f_c - f_n$ to the carrier amplitude at $f_c$  is directly related to $\zeta_n$ as $\zeta_n = \frac{\mu}{2 \pi/\lambda}$. With experimentally measured sidebands and carrier amplitudes, we use $\lambda$ to calibrate the oscillation amplitude of the resonator in optical measurements \cite{wagner1990optical}.

Connecting $\rm PD_2$ directly to a spectrum analyzer allows for  measuring the thermal noise of the NEMS, which is used to determine the spring constants  of its different modes. Connecting $\rm PD_2$ to a lock-in amplifier (SR844, Stanford Research Systems) allows for measuring the resonant response of the NEMS. The block diagram for measuring the driven response is shown in Fig. \ref{Figure Optsetup}(b).

\subsection{Device Fabrication and Dimensions}

The strain gauges and contact pads are patterned using electron beam lithography (EBL). A 135-nm-thick gold film and a 5-nm chromium adhesion layer underneath are deposited by thermal evaporation. The vacuum during the deposition is typically less than $1 \times 10^{-5}$ mBar. The deposition rate is less than 0.1 nm/s. The estimated substrate temperature during deposition is $\approx 50^{\rm o} \rm C$. The sample is cooled inside the vacuum after the deposition before exposing to atmosphere.  After lift-off, a second step of EBL is performed to define the beam structures. A 60-nm-thick copper film is deposited via electron beam evaporation, which serves as a dry etch mask. Inductively coupled plasma etching is used to etch the silicon nitride layer anisotropically; then, the beam is suspended by an isotropic etch step that removes the silicon beneath. After the removal of the copper etch mask, the devices are placed in a chip carrier with 50-$\Omega$ strip lines and wirebonds are made between the contact pads and the striplines.

\begin{figure}
    \centering
    \includegraphics[width= 4.0in]{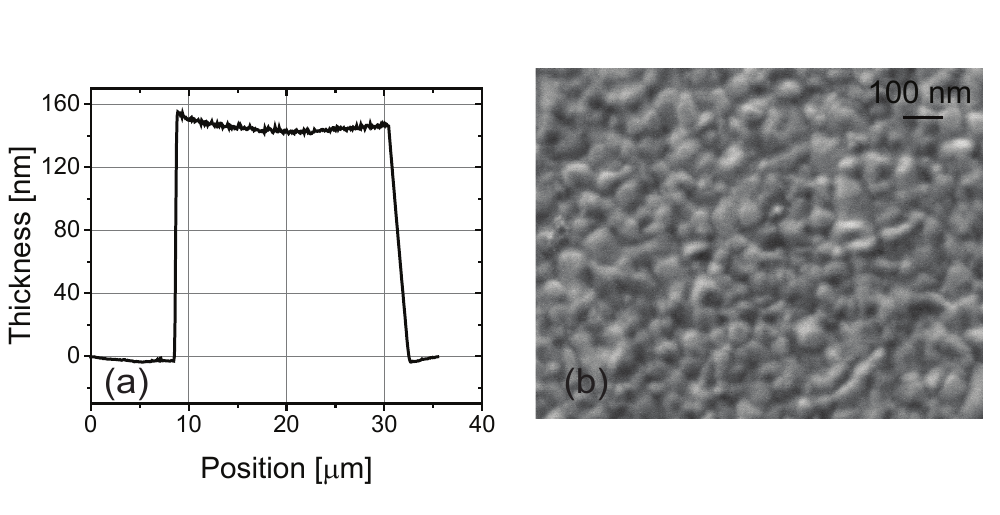}
    \caption{(a) Atomic Force Microscope (AFM) measurement of the thickness  of the gold thin film. (b) Scanning electron microscope (SEM) image of the grains in the film.}
    \label{Figure grain}
\end{figure} 

\subsection{Physical Properties of the Silicon Nitride and the Gold}\label{Physical Properties}

The density and Young's modulus for silicon nitride is reported to be in the range $2600~{\rm kg/m^3} \le \rho \le 3400 ~{\rm kg/m^3}$ and $250 ~{\rm GPa} \le E \le 350~{\rm GPa}$, respectively \cite{tabata1989mechanical}. We use X-ray reflectometry (XRR) to obtain the density as $2960 \pm 30~{\rm kg/m^3}$. With the measured eigen-mode frequencies of the resonator, we estimate that the tension is $S\approx 68.8 \pm 12.9~ \rm \mu N$ \cite{ari2020nanomechanical}. We determine the thickness and the grain size of the gold thin-film from Atomic Force Microscopy (AFM) and Scanning Electron Microscopy (SEM), as shown in Fig. \ref{Figure grain}.  AFM measurements indicate that the thickness of the film is $144 \pm 5 \rm ~nm$, consistent with the deposition thickness during fabrication. The SEM image shown in Fig. \ref{Figure grain}(b) has provided an estimate of the grain size to be $55 \pm \rm 40 ~ nm$.  

\section{Optical Measurements}
\subsection{Measurement of Each Eigen-Mode}\label{Eigen-Mode}

We use  point measurements at the anti-node positions to measure the  eigen-modes of the NEMS resonator. We assume that each mode is a lumped single-degree-of-freedom system independent from the other modes due to the high $Q$ factors.  We describe the  resonator mode in terms of its oscillation amplitude $z$ at the anti-node position  as a mass-spring system with effective mode mass $m_n$ and  spring constant $k_n$ \cite{ari2020nanomechanical}. When the beam is exposed to  a drive force $F(t)=\sqrt{2}F_n \cos2\pi ft$ with rms amplitude $F_n$ and frequency $f\approx f_n$, the mechanical motion of the mode can be expressed as
\begin{equation}
    m_n \ddot{z} + m_n \gamma_n \dot{z} + k_n z = F(t),
    \label{eq S1}
\end{equation}
where $k_n =  m_n (2 \pi f_n)^2$ and $\gamma_n$ are the effective spring constant and the intrinsic damping coefficient of the beam in its $n^{\rm th}$ mode, respectively. Here, $ {f_n}$ is the eigen-mode frequency.  This drive generates a response $z(t)=\sqrt{2}\zeta_n(f) \cos \big[2\pi ft + \phi(f)\big]$. After some manipulations \cite{schmid2016fundamentals}, one  obtains the following expressions for the rms  amplitude of the beam in its $n^{\rm th}$   mode 
\begin{equation}
    {\zeta_n (f)} =\frac{F_n}{\sqrt{{\left (k_n - 4 \pi^2 m_n f^2\right)^2 + \left(4 \pi^2 {m_n}  f \gamma_n\right)^2}}} = \frac{F_n/m_n}{\left(2 \pi\right)^2\sqrt{ \left( {f_n}^2 - f^2\right)^2 + \left(\frac{f_n f}{Q_n} \right)^2}}
    \label{eq S2},
\end{equation}
where  $Q_n = \frac{2 \pi f_n}{\gamma_n}$. If the beam mode is just driven by thermal fluctuations, then the power spectral density of the displacement fluctuations of the mode is
\begin{equation}
    G_{n} (f) = \frac{4k_B T  f_n}{\left(2 \pi\right)^4 m_n Q_n \bigg[\left( {f_n}^2 - f^2\right)^2 + \left(\frac{f_n f}{Q_n} \right)^2\bigg]}.
    \label{eq S3}
\end{equation}

\begin{figure}
    \centering
    \includegraphics[width= 5.8in]{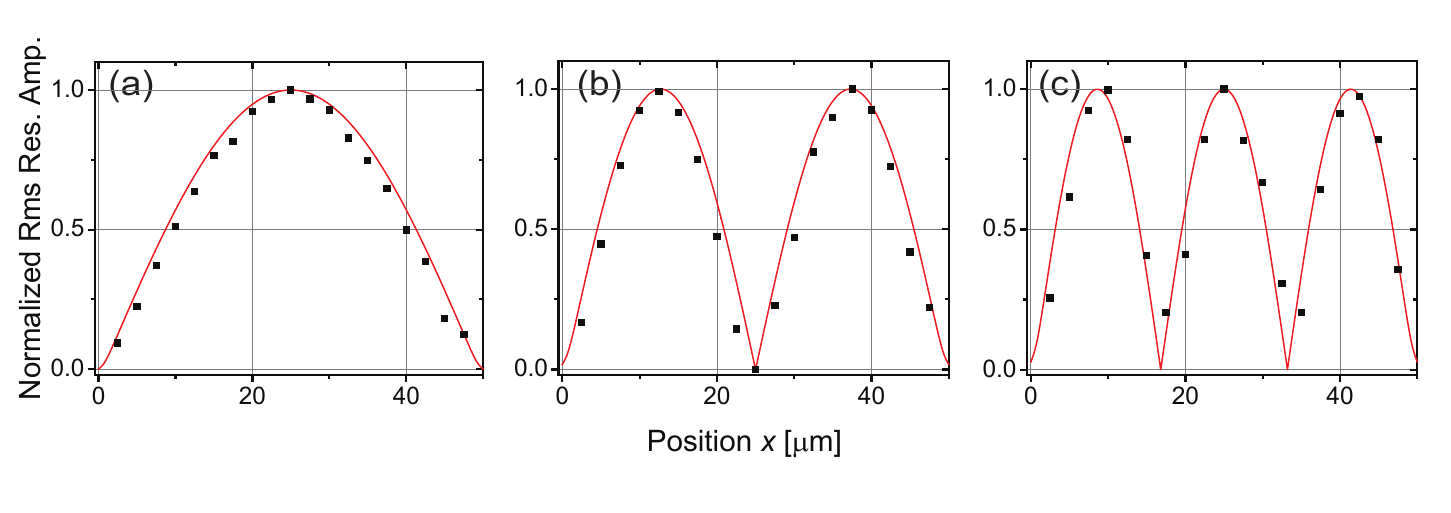}
    \caption{(a)-(c) The first three eigen-mode shapes of a 50-$\rm \mu$m-long beam resonator. The black squares show the rms amplitude of the beam  as the laser spot is scanned along the length of a 50-$\rm \mu$m-long beam, i.e., $x$ axis, at a fixed drive voltage. The red curves are the modes shapes obtained from a COMSOL simulation.}
    \label{Figure Mode shape}
\end{figure}

To precisely determine the anti-node positions of the NEMS resonator, we measure the mode shapes of the beam by scanning the laser spot along the length of the resonator (i.e., $x$ axis shown in Fig. 1 in the main text or \ref{Figure Model}). Figs. \ref{Figure Mode shape}(a-c) show the normalized rms resonance amplitudes for the first three modes of a 50-$\mu$m-long resonator as a function of $x$.  Here, the drive voltage for all modes is kept constant. To obtain each data point (the black square), we drive the resonator around its eigen-mode frequency and measure the resonance curve. We construct the data traces by taking the peak value from each resonance curve, obtaining the mode shape.  The measured mode shapes match well with those predicted in COMSOL simulations  (Fig. \ref{Figure Mode shape} solid curves). We can thus position the optical spot on the anti-node positions very accurately with a typical error  $ \lesssim 1~\mu \rm m$ in the measurements.  The error is further reduced  by maximizing the resonance amplitude around the anti-node. In the end, we estimate the maximum error in resonator amplitude due to uncertainty in the anti-node position to be $\lesssim 2\%$, which is taken as a source of error in Section \ref{Error Analysis}.

Figs. \ref{Figure Opt30}, \ref{Figure Opt50}, and \ref{Figure Opt60} respectively show the oscillation amplitude $\zeta_n$ as a function of frequency at different driving voltages for the first four eigen-modes of a 30-$\rm \mu$m-long, a 50-$\rm \mu$m-long, and a 60-$\rm \mu$m-long resonator. Here, the beam is driven electrothermally by applying a harmonic voltage at different  amplitudes, with the frequency swept around each eigen-mode frequency. Figs. \ref{Figure Opt30_thermal}, \ref{Figure Opt50_thermal}, and \ref{Figure Opt60_thermal} show the PSDs of the thermal fluctuations of the corresponding modes. Using the PSD for each mode, we determine effective mode spring constant $k_n$ of the beam (see below). 
We note that the thermal noise and  driven measurements are performed on  different beams with the same nominal dimensions from the same batch. This accounts for the slight differences in mode frequencies obtained in thermal and driven experiments --- as commonly encountered in the NEMS literature \cite{ekinci2005nanoelectromechanical}. Also, the $Q$ factors for the same mode will be different when measured at (slightly) different pressures.  In summary, from these measurements of each eigen-mode, we  determine the eigenfrequency $f_n$, spring constant $k_n$, and quality factor $Q_n$ for each mode.  

\subsection{Amplitude \textit{vs.} Drive and Strain \textit{vs.} Amplitude for Each Mode}\label{strain calibration}

\begin{figure}
    \centering
    \includegraphics[width= 3.9in]{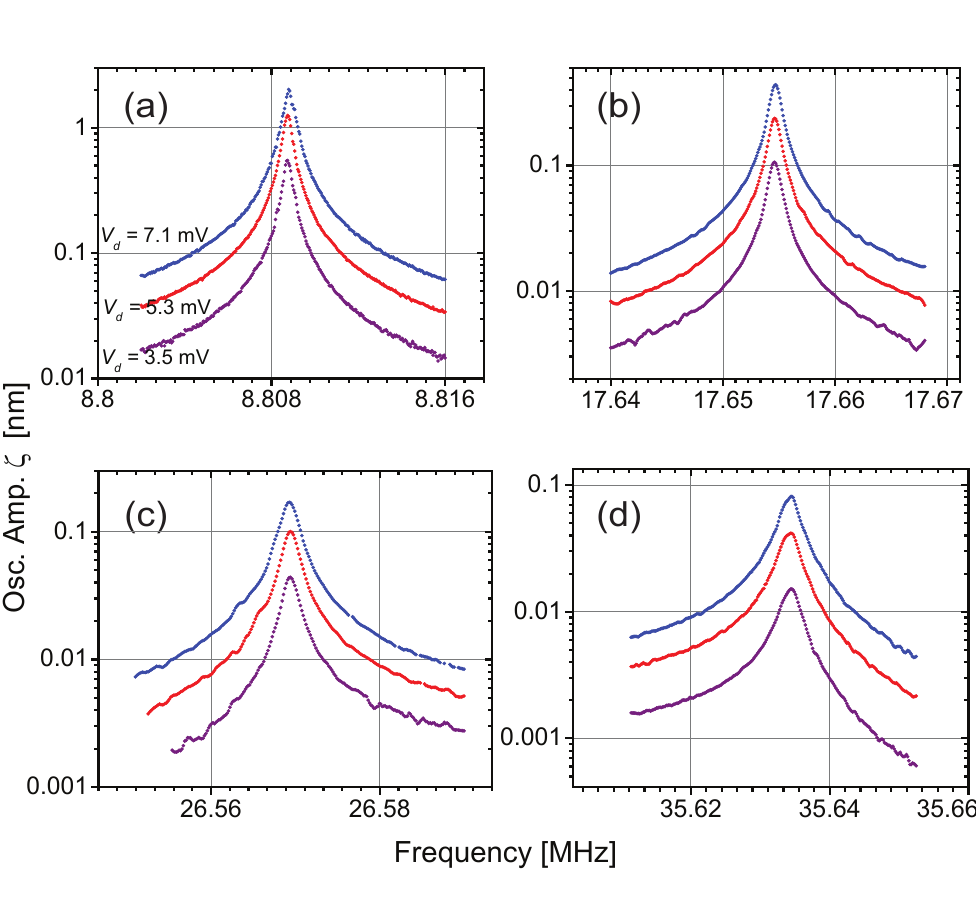}
    \caption{(a)-(d) Rms oscillation amplitudes $\zeta_n$ of a 30-$\rm \mu$m-long beam as a function of frequency around its first four eigen-mode frequencies measured at the anti-node positions plotted in a semi-logarithmic plot. The rms drive voltages applied to the beam are  7.1  mV,  5.3  mV,  and  3.5  mV  for  the  blue,  red,  and  brown  curves, respectively.}
    \label{Figure Opt30}
\end{figure}

\begin{figure}
    \centering
    \includegraphics[width= 3.9in]{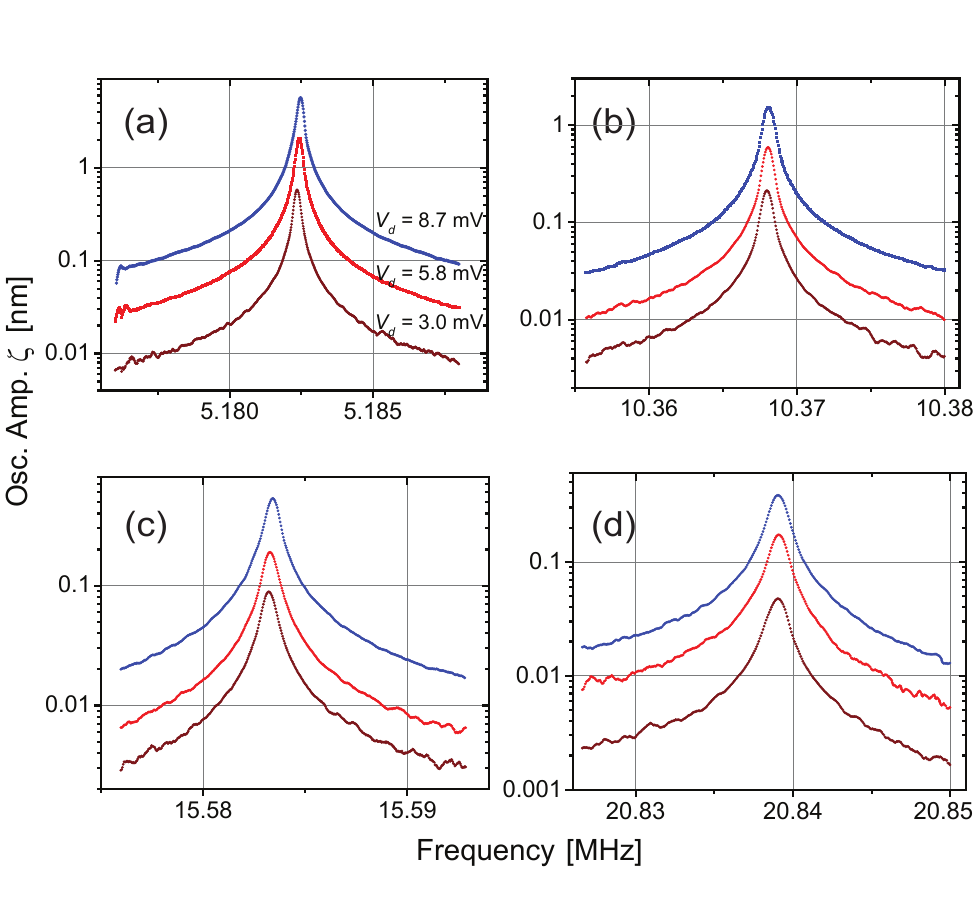}
    \caption{(a)-(d) Rms oscillation amplitudes $\zeta_n$ of a 50-$\rm \mu$m-long beam as a function of frequency around its first four eigen-mode frequencies measured at the anti-node positions plotted in a semi-logarithmic plot. The rms drive voltages applied to the beam are  8.7  mV,  5.8  mV,  and  3.0  mV  for  the  blue,  red,  and  brown  curves, respectively.}
    \label{Figure Opt50}
\end{figure}

\begin{figure}
    \centering
    \includegraphics[width= 3.9in]{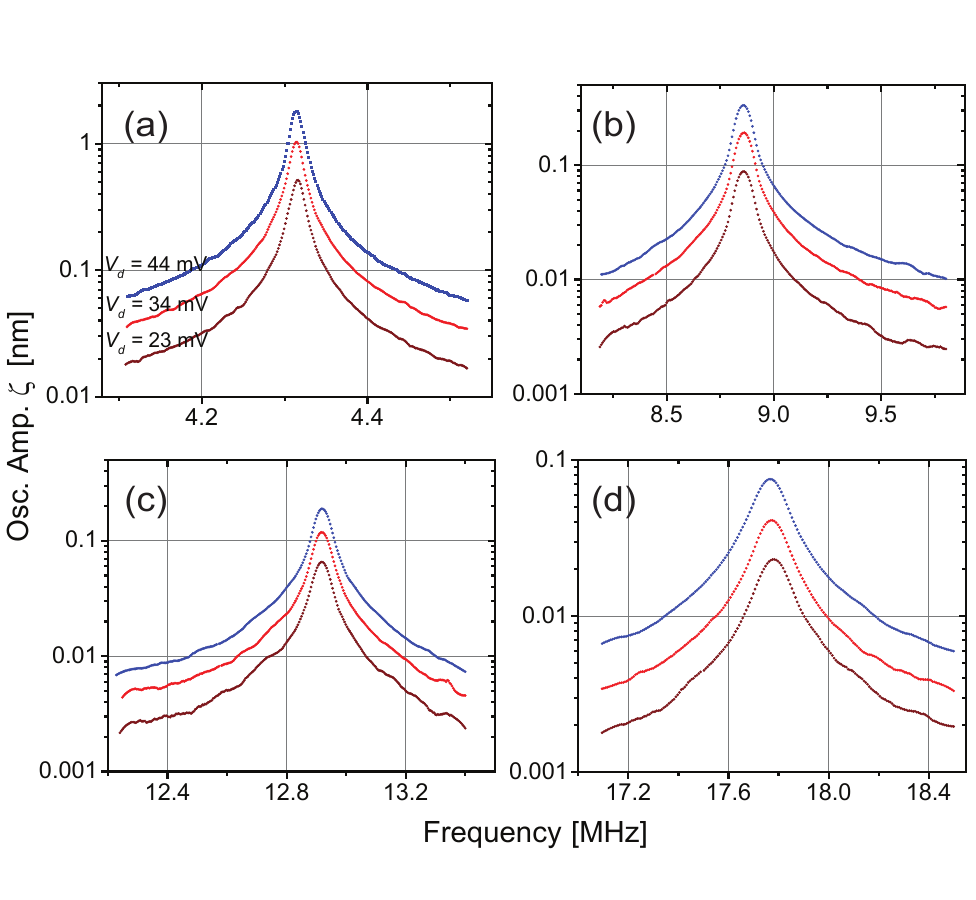}
    \caption{(a)-(d) Rms oscillation amplitudes $\zeta_n$ of a 60-$\rm \mu$m-long beam at different drives around its first four eigen-mode frequencies plotted in a semi-logarithmic plot. The measurements are performed at the anti-node of each mode. The rms drive voltages applied to the beam are  44  mV,  34  mV,  and  23  mV  for  the  blue,  red,  and  brown  curves, respectively.}
    \label{Figure Opt60}
\end{figure}

\begin{figure}
    \centering
    \includegraphics[width= 6.75in]{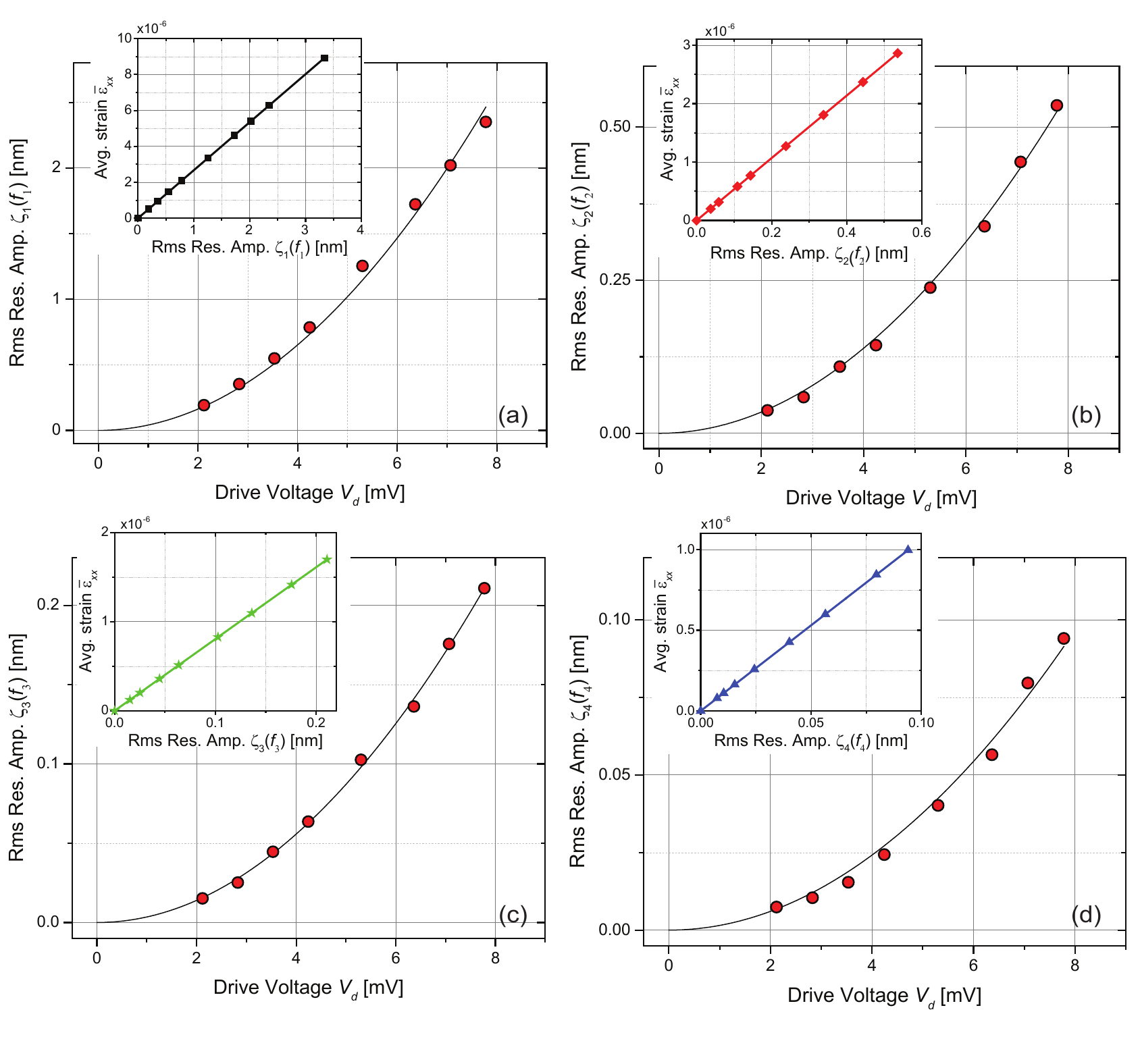}
    \caption{(a)-(d) Rms resonance amplitudes for the first four eigen-modes of a 30-$\rm \mu m$-long beam at different drive voltages. The corresponding insets in (a)-(d) show the calculated strains on the strain gauge as a function of the rms resonance amplitude at each mode.}
    \label{Figure strainVSAmp_30um}
\end{figure}

\begin{figure}
    \centering
    \includegraphics[width= 6.75in]{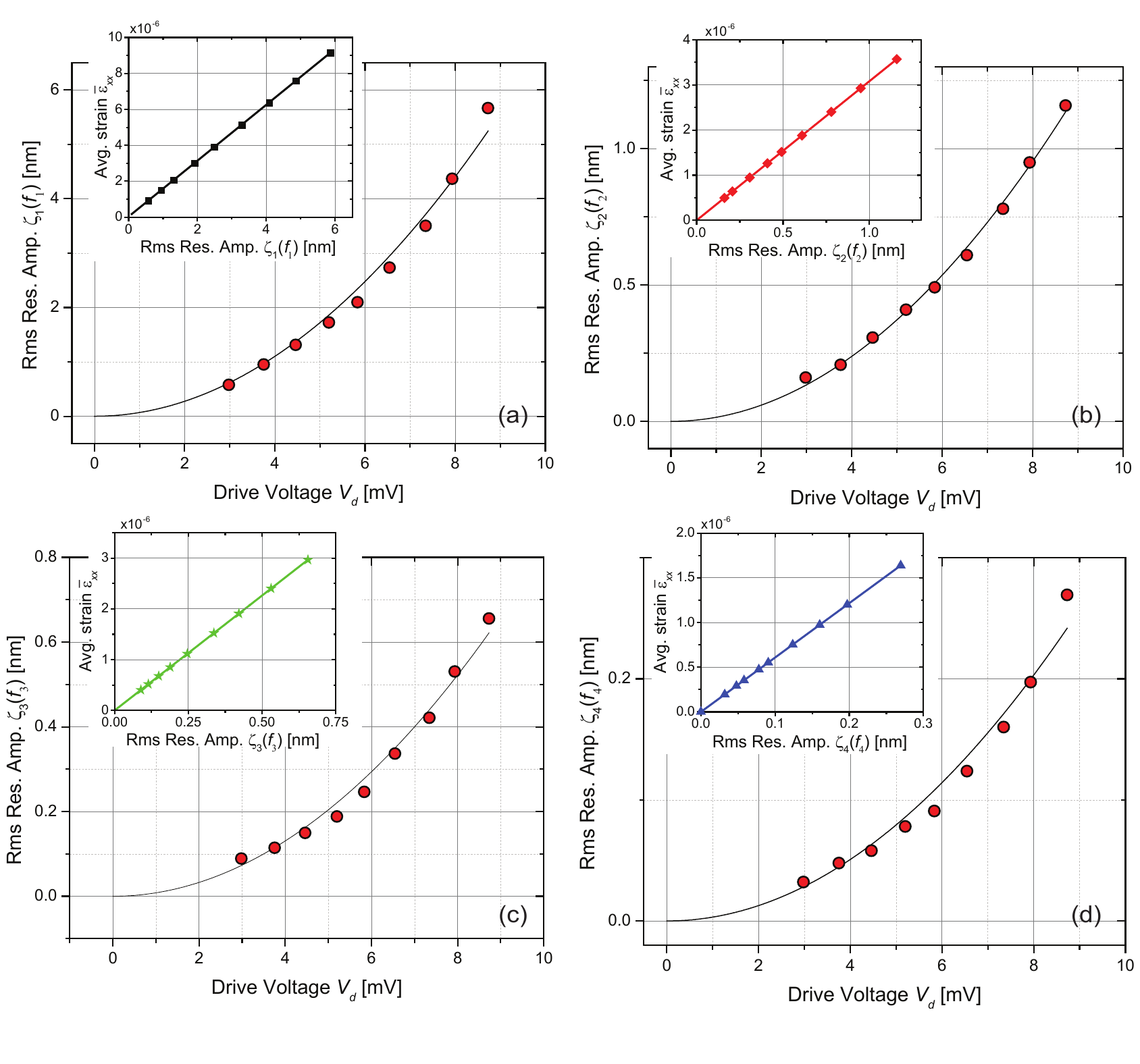}
    \caption{(a)-(d) Rms resonance amplitudes for the first four eigen-modes of a 50-$\rm \mu m$-long beam at different drive voltages. The corresponding insets in (a)-(d) show the calculated strains on the strain gauge as a function of the rms resonance amplitude at each mode.}
    \label{Figure strainVSAmp}
\end{figure}

\begin{figure}
    \centering
    \includegraphics[width= 6.75in]{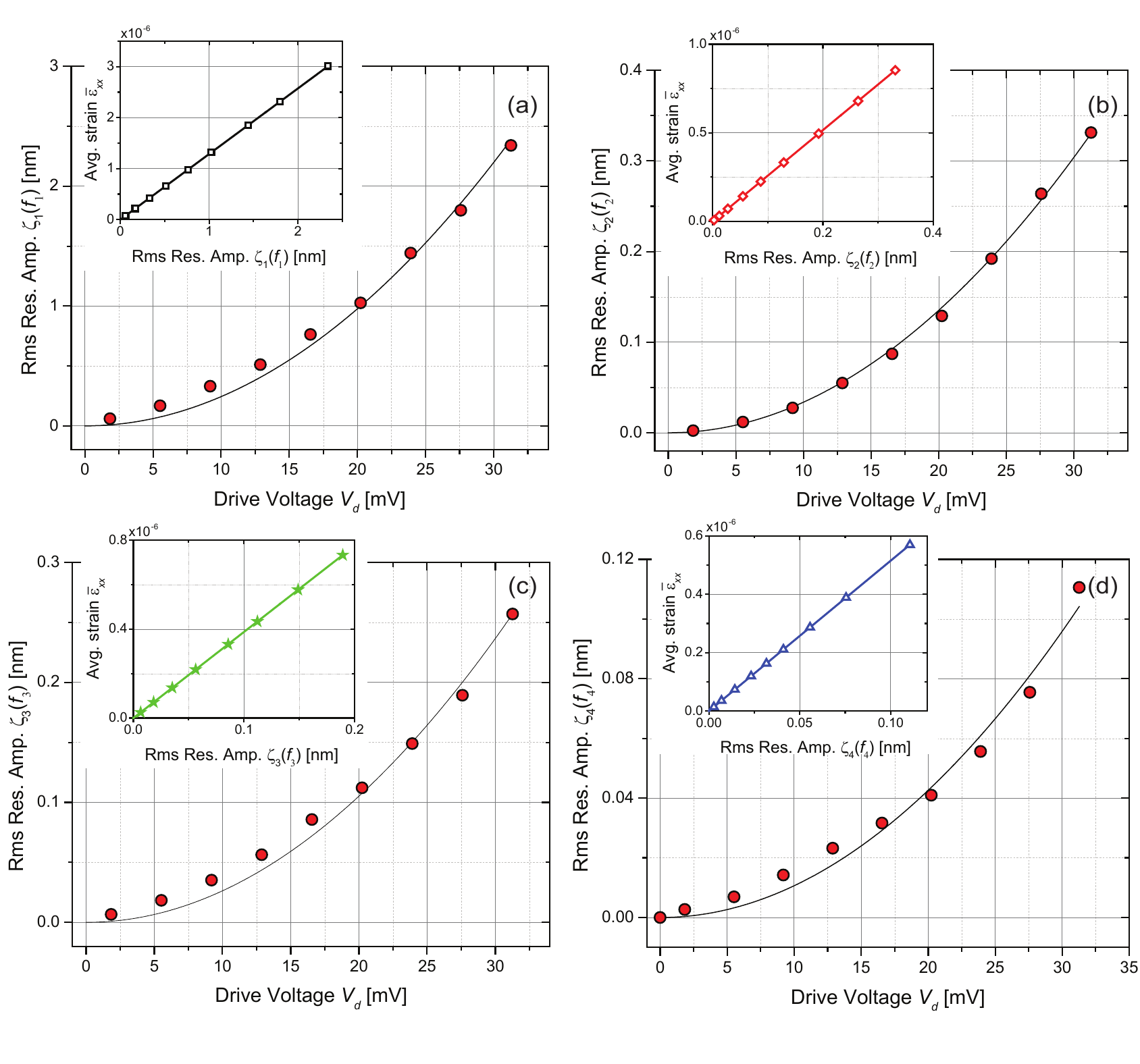}
    \caption{(a)-(d) Rms resonance amplitude for the first four eigen-modes of a 60-$\rm \mu m$-long beam as a function of the drive voltage. The insets in (a)-(d) show the calculated strains on the strain gauge as a function of the rms resonance amplitude at each mode.}
    \label{Figure strainVSAmp_60um}
\end{figure}

In this section, we describe how we find the strain for the first four modes of the 30-$\rm \mu$m-long,  50-$\rm \mu$m-long, and  60 $\rm \mu$m-long resonators. Figs. \ref{Figure Opt30}, \ref{Figure Opt50} and \ref{Figure Opt60} show the  resonance curves for these beams. The data  in Figs. \ref{Figure strainVSAmp_30um}, \ref{Figure strainVSAmp}, and \ref{Figure strainVSAmp_60um}(a-d) are taken from these resonance curves and show the peak values, i.e., the rms resonance amplitudes $\zeta_n(f_n)$, for each mode as a function of the applied drive voltages $V_d$.  

\begin{figure}
    \centering
    \includegraphics[width= 2.3in]{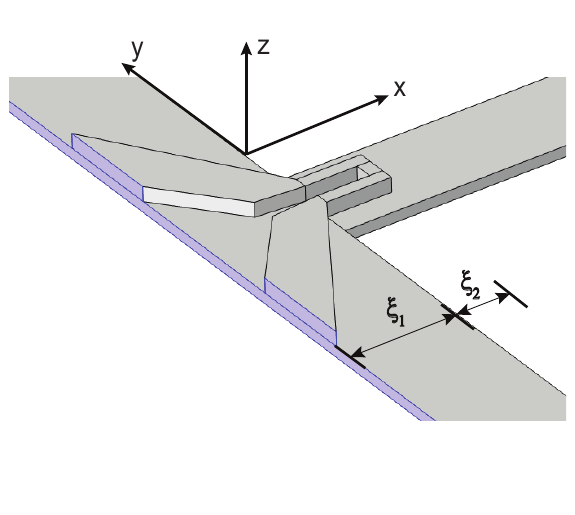}
    \caption{Device model used in the simulations. It includes the strain gauge (u-shaped nanoresistor) and the silicon nitride resonator with the undercut. The nanoresistor strain gauge on top of the undercut region ($\xi_1 \approx 800$ nm) and the beam ($\xi_2 \approx 600$ nm)  are both exposed to strain when the beam vibrates in one of its eigen-modes. The blue surface shows one of the fixed boundaries.}
    \label{Figure Model}
\end{figure}

We solve for the strain field $ \varepsilon_{xx} (\mathbf{r})$ on the strain gauge using  the finite element method (FEM) in a commercial software, COMSOL \cite{comsol}. We implement  a stationary study and a pre-stressed eigenfrequency study in the Solid Mechanics Module. The model of the device includes the gold strain gauge and the silicon nitride resonator including the undercut, as shown in Fig. \ref{Figure Model}. The dimensions in the model are taken from SEM and AFM measurements as well as from fabrication parameters.  The resonator is fixed at  both ends, with one of the fixed boundaries shown in Fig. \ref{Figure Model} as the blue surface. As can be seen, the undercut region is also a part of the resonator. The undercut region extends ~800 nm in the $x$ direction. To model the tension in the beams, we impose an initial stress. The magnitude of the stress is determined by minimizing the error between the experimentally-measured and the numerically-solved eigenfrequencies of each beam. From the simulation, we find the stress to be $\sigma = $ 695 MPa, which results in an axial tensile load of $S \approx 65~\rm \mu$N.  The value of the axial load also agrees closely with analytical  estimations (see Section \ref{Physical Properties} and \cite{ari2020nanomechanical}). For each mode, the simulation provides the  mode shape (red curves in Figs. \ref{Figure Mode shape}(a-c)).  We impose an rms displacement amplitude  over the entire beam such that the rms displacement amplitude is $\zeta_n (f_n)$ at the anti-nodes. From the ensuing deformation, we obtain the strain field $ \varepsilon_{xx} (\mathbf{r})$ on the strain gauge. We then compute the average strain $\bar \varepsilon_{xx}$ as discussed in the main text. The insets in Figs. \ref{Figure strainVSAmp_30um}(a-d), Figs. \ref{Figure strainVSAmp}(a-d), and Figs. \ref{Figure strainVSAmp_60um}(a-d) respectively show the average strain $\bar \varepsilon_{xx}$ on the strain gauge as a function of $\zeta_n (f_n)$ for the first four modes of the 30-$\rm \mu$m-long, 50-$\rm \mu$m-long, and  60-$\rm \mu$m-long resonators. 

\subsection{Eigenfrequencies, Spring Constants, and Quality Factors} \label{freq spring Q}

Due to the stress in the material, the eigenfrequencies $f_n$ of the resonator are dominated by  tension, $f_n \propto n$ \cite{ari2020nanomechanical, liem2020inverse}. Figs. \ref{Figure Freq_k_Q_30}(a), \ref{Figure Freq_k_Q}(a), and \ref{Figure Freq_k_Q_60}(a) show the first four  $f_n$ for the 30-$\rm \mu$m-long,  50-$\rm \mu$m-long, and  60-$\rm \mu$m-long resonators as a function of the mode number $n$, respectively. 

\begin{figure}[b]
    \centering
    \includegraphics[width= 3.9in]{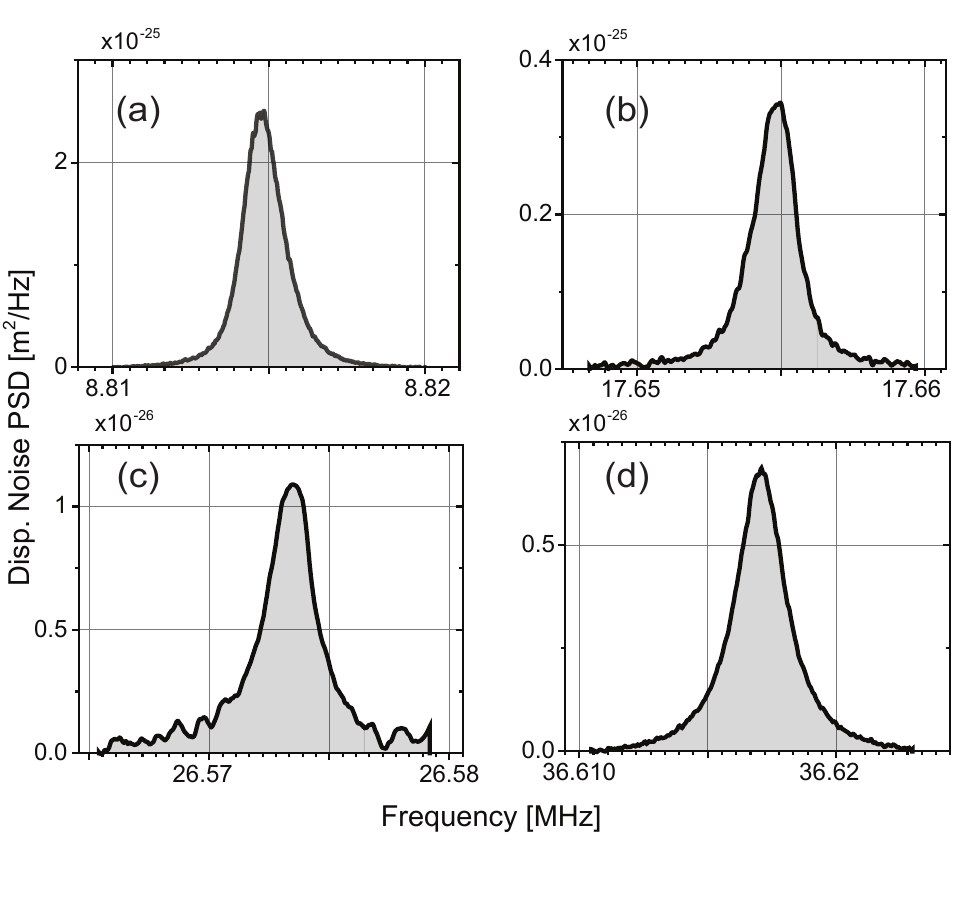}
    \caption{(a)-(d) Power spectral densities (PSDs) of the equilibrium displacement  fluctuations of the first four eigen-modes of a 30-$\rm \mu$m-long resonator. The measurements are performed at the anti-node position of each mode.}
    \label{Figure Opt30_thermal}
\end{figure}

\begin{figure}
    \centering
    \includegraphics[width= 3.9in]{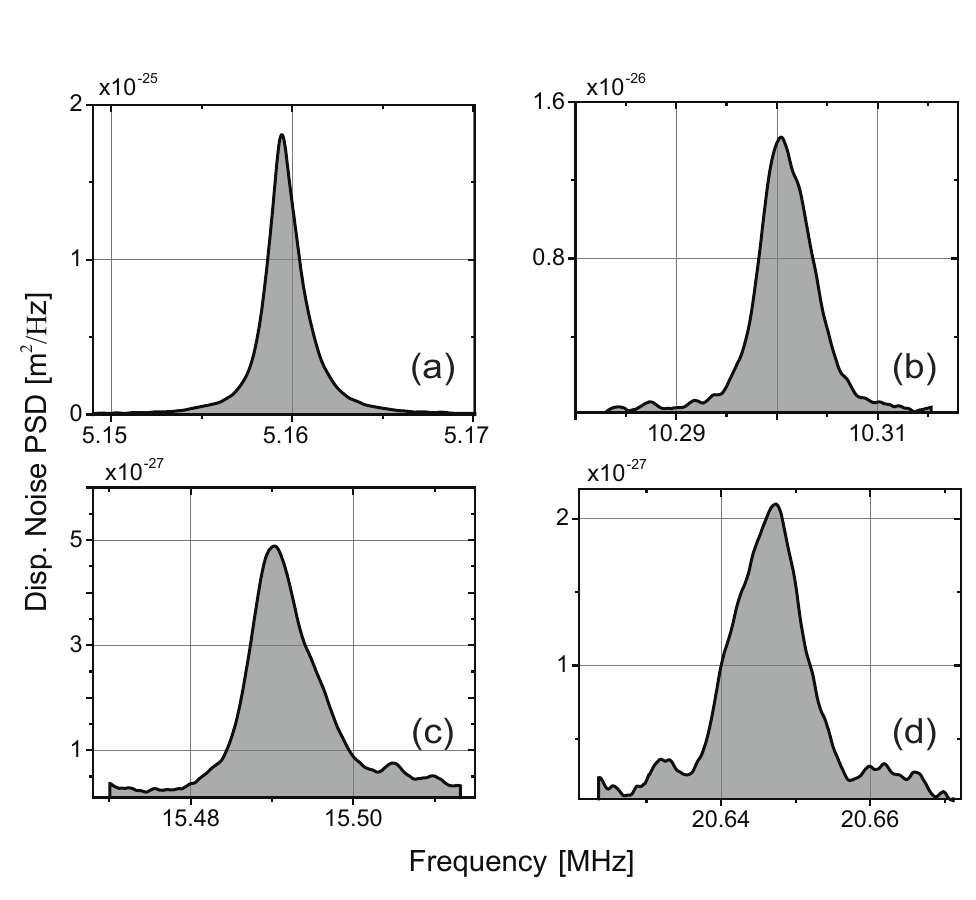}
    \caption{(a)-(d) Power spectral densities (PSDs) of the equilibrium displacement  fluctuations of the first four eigen-modes of a 50-$\rm \mu$m-long resonator. The measurements are performed at the anti-node position of each mode.}
    \label{Figure Opt50_thermal}
\end{figure}

\begin{figure}
    \centering
    \includegraphics[width= 3.9in]{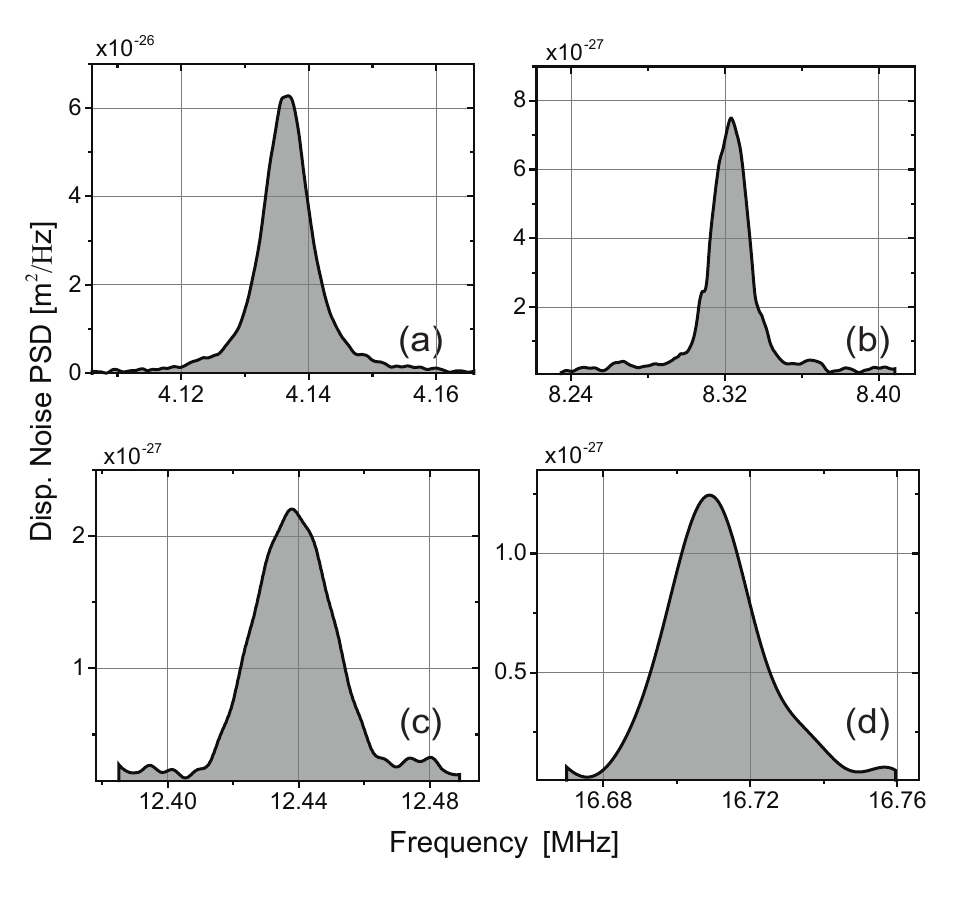}
    \caption{(a)-(d) Power spectral densities (PSDs) of the equilibrium displacement  fluctuations of the first four eigen-modes of a 60-$\rm \mu$m-long resonator. The measurements are performed at the anti-node  position of each mode.}
    \label{Figure Opt60_thermal}
\end{figure}

\begin{figure}
    \centering
    \includegraphics[width= 5.8in]{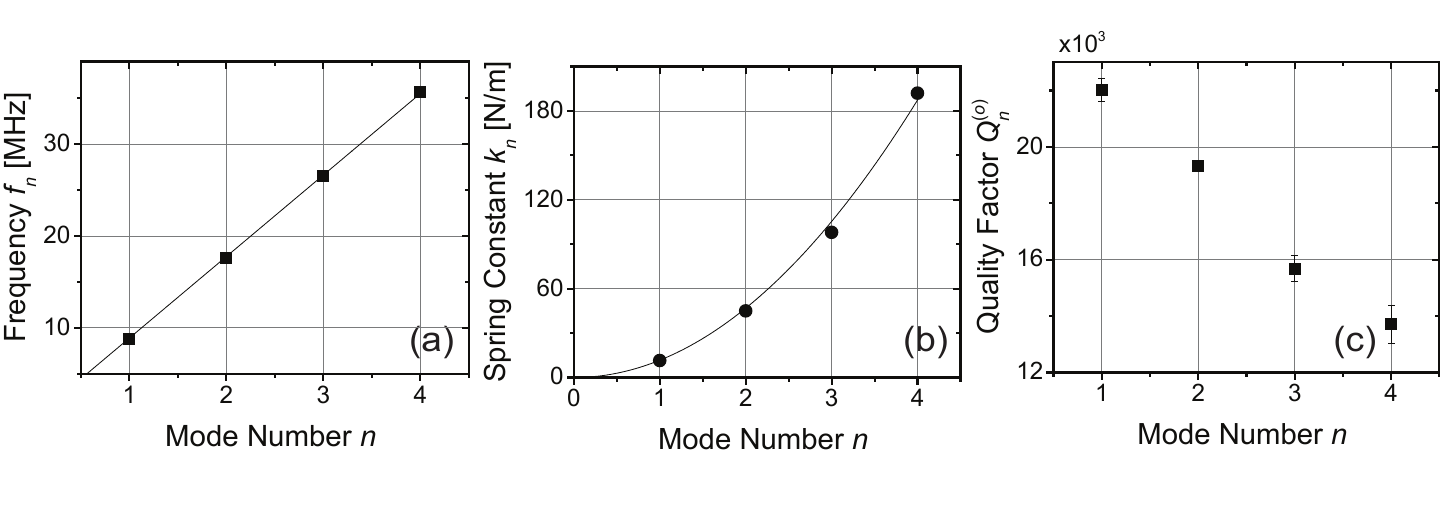}
    \caption{(a)-(c) Eigenfrequencies $f_n$, spring constants $k_n$, and  quality factors $Q_n^{({o})}$ of the first four eigen-modes of the 30-$\rm \mu$m-long resonator. 
    The error bar for each mode in (c) shows the standard deviation of $Q$ measured at different drive voltages (amplitudes).}
    \label{Figure Freq_k_Q_30}
\end{figure}

\begin{figure}
    \centering
    \includegraphics[width= 5.8in]{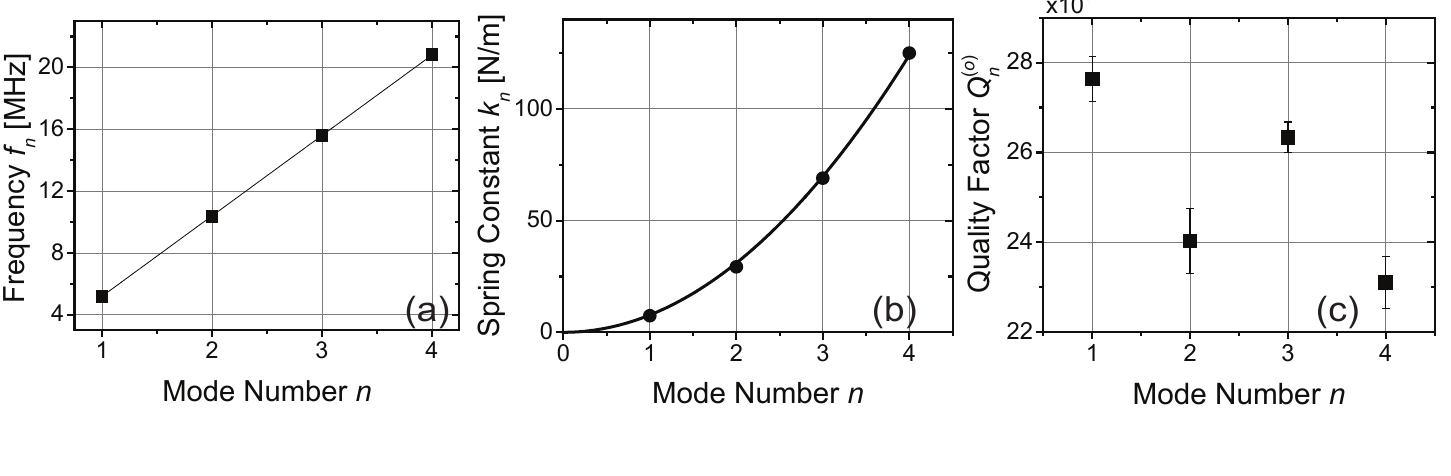}
    \caption{(a)-(c) Eigenfrequencies $f_n$, spring constants $k_n$, and  quality factors $Q_n^{({o})}$ of the first four eigen-modes of the 50-$\rm \mu$m-long resonator. 
    The error bar for each mode in (c) shows the standard deviation of $Q$ measured at different drive voltages (amplitudes). }
    \label{Figure Freq_k_Q}
\end{figure}

\begin{figure}
    \centering
    \includegraphics[width= 5.8in]{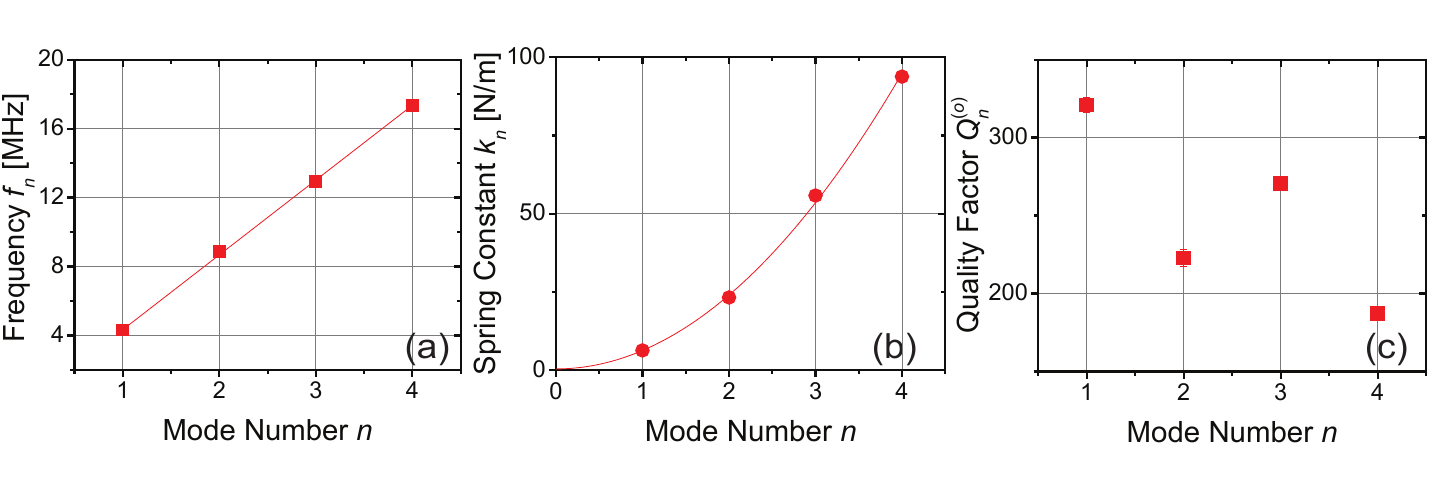}
    \caption{(a)-(c) Eigenfrequencies $f_n$, spring constants $k_n$ and the   quality factors $Q_n^{({o})}$ of the first four eigen-modes of the 60-$\rm \mu$m-long resonator. The error bar for each mode in (c) is the standard deviation of $Q$ measured at different drive voltages (amplitudes). The measurements of $Q$ here are performed at a higher pressure compared to the data in Figs.~\ref{Figure Freq_k_Q}(c) and \ref{Figure Freq_k_Q_30}(c).}
    \label{Figure Freq_k_Q_60}
\end{figure}

We determine the the spring constants $k_n$  from the classical equipartition theorem, $k_n = \frac{k_B T}{\langle {x_n}^2\rangle}$, where $k_B$ is the Boltzmann constant, $T$ is the temperature, and $\langle {x_n}^2\rangle$ is  mean-squared thermal amplitude for the $n^{\rm th}$ mode at an anti-node.  Figs. \ref{Figure Opt30_thermal}, \ref{Figure Opt50_thermal}, and \ref{Figure Opt60_thermal}(a-d) show the power spectral densities (PSDs) of the first four eigen-modes of  the 30-$\rm \mu$m-long,  50-$\rm \mu$m-long, and  60-$\rm \mu$m-long resonators, respectively. Since  $Q$ is high, we calculate  $\langle {x_n}^2\rangle$ from the integral of the PSD (the shaded areas in Figs. \ref{Figure Opt30_thermal}, \ref{Figure Opt50_thermal}, and \ref{Figure Opt60_thermal}).  Figs. \ref{Figure Freq_k_Q_30}(b), \ref{Figure Freq_k_Q}(b), and \ref{Figure Freq_k_Q_60}(b) show the spring constants $k_n$ for the  resonators as a function of the mode number $n$. Numerical values of  $k_n$ for both resonators are also listed in  Table $\rm I$ in the main text. The spring constants of all NEMS resonators are quadratically dependent on  mode number. Since the effective mode mass $m_n$ is constant \cite{ari2020nanomechanical} for different modes of the same beam and $f_n$ is linearly dependent on $n$, $k_n$ is proportional to ${f_n}^2$.

From both optical and electrical measurements, we obtain $Q$ factors of the modes of the beam, $Q_n^{({o})}$ and $Q_n^{(e)}$, respectively, by fitting the driven resonance peak of the NEMS resonator to a Lorentzian (i.e, Eq. \ref{eq S2}). Figs. \ref{Figure Freq_k_Q_30}(c), \ref{Figure Freq_k_Q}(c), and \ref{Figure Freq_k_Q_60}(c) show the $Q_n^{(o)}$ as a function of mode number for the 30-$\rm \mu m$-long,  50-$\rm \mu m$-long, and  60-$\rm \mu m$-long NEMS resonators, respectively. Below, in  Figs. \ref{Figure Electr_30}(e), \ref{Figure Electr_50}(e), and \ref{Figure Electr_60}(e), we show  $Q_n^{(e)}$ for the first four eigen-modes for the same resonators. {The error bar at each mode is calculated from  the slightly different  $Q$ factors measured at different drive amplitudes.}

We note that both $Q_n^{(o)}$ and $Q_n^{(e)}$ for the same mode of the same resonator can vary from measurement to measurement due to, e.g., chamber pressure, resonator surface conditions, and so on.  For a given drive voltage $V_d$ and  force $F_n$, the oscillation amplitude $\zeta_n(f_n)$ depends  on the quality factor, i.e., $\zeta_n(f_n) = \frac{F_n Q_n^{(o)}}{k_n}$ or $\zeta_n(f_n) = \frac{F_n Q_n^{(e)}}{k_n}$, because $k_n$ is a constant and the measured electrical or optical signal is linearly proportional to $\zeta_n(f_n)$. In order to compensate for small variations in the $Q$ factors from measurement to measurement, we scale the resonance amplitude with the $Q$ factor appropriately --- as described in the main text.  

\clearpage

\section{Electrical Measurements}

\subsection{Resisitivity of the Gold Film Electrodes}

\begin{figure}[b]
    \centering
     \includegraphics[width=6.75in]{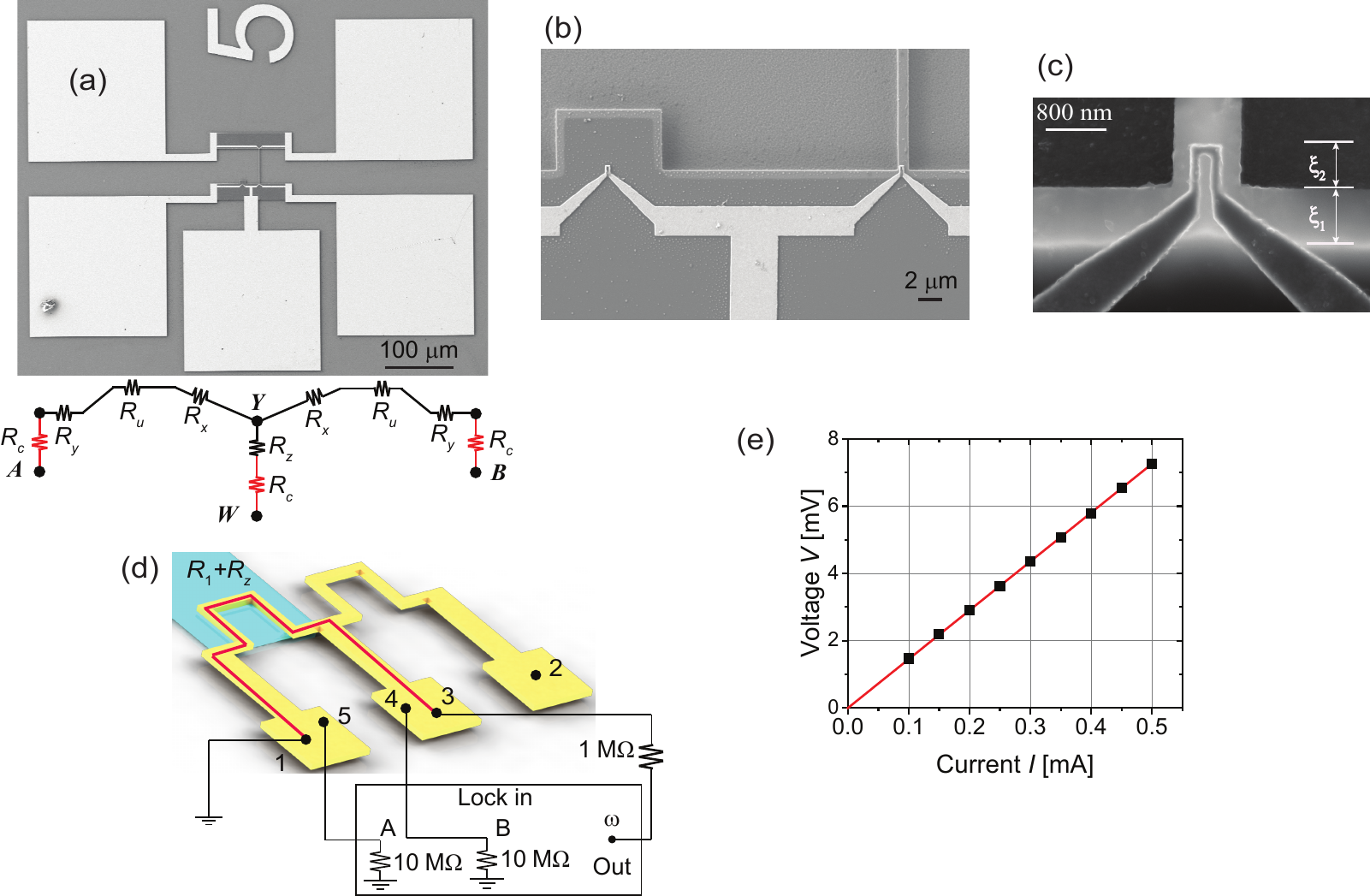}
    \caption{(a) SEM images of the entire device (top) and the electrical resistance model. (b) The nanoresistors and the connecting electrical paths. (c) A close up of a nanoresistor.   The undercut region is $\xi_1 \approx$ 800 nm, and the region on the resonator is $\xi_2 \approx$ 600 nm.  (d) Schematic diagram for the 4-wire resistance measurement. The lock-in amplifier (SR830, Stanford Research Systems) and the $1~\rm M \Omega$ source resistor provide a current  at $  1~ \rm kHz$. The voltage drop on $R_1+R_{z}$, where $R_1 = R_{x}+ R_{u} + R_{y}$, is detected using the $A-B$ mode of the lock-in.  (e) Measured voltage (symbols) at different currents. The red line is a linear fit  providing $R_1+R_{z} = 14.51~ \Omega$. }
    \label{Figure sheetRes}
\end{figure}

Fig. \ref{Figure sheetRes}(a-c) shows SEM images of the  electrode patterns and the strain gauge. The resistors in the circuit diagrams  can be traced to the SEM images. The total resistance of the electrode is determined from a four-wire measurement.  In the four-wire measurement shown in Fig. \ref{Figure sheetRes}(d), two separate wire bonds are attached to each bonding pad.  The 1 $\rm M \Omega$ source resistor turns the lock-in reference voltage output into a current source. The voltage drop across $R_{y}+ R_{u} + R_{x} + R_{z}$ is monitored in the $A-B$ detection mode of the lock-in amplifier (SR830, Stanford Research Systems). Fig. \ref{Figure sheetRes}(e) shows the $IV$ curve from the measurement, which yields a resistance of $14.51~ \Omega$.

To find the resistivity, we integrate the electrode shape along the  path of current flow, with the differential resistance of the infinitesimal length $d \ell$ along the path  being $dR=\frac{\rho d\ell}{ W(\ell)h}$. We emphasize that the area through which current flows, $ W(\ell) h$, is position dependent. We equate the result of this integration to the measured resistance value and find $\rho=2.81 \times 10^{-8}~ \rm \Omega \cdot m$. In an alternative approximation, we ``count the number of squares" along the current path. Most of the electrode has  rectangular geometry and squares can be formed easily, with the corner squares (where the current takes a turn) counted as 0.56 squares \cite{jaeger2002introduction}. The region leading to the nanoresistors has a  more complicated shape and is approximated in units of squares along the current path. We find the sheet resistance to be  $0.24 \pm 0.01 ~ {\Omega}/{\square}$, which yields $\rho=2.83 \times 10^{-8}~ \rm \Omega \cdot m$. That these two values are quite close gives us confidence. In the calculations, we use the average of these two values.

Using $\rho$ and the geometry for each region, we calculate the resistance values for $R_{u}$, $R_{x}$, $R_y$ and $R_z$, as reported  in Table I in the main text. The resistances for the wirebonds and contacts are determined from two-wire resistance measurements. Here, two-wire resistances are measured in different combinations, and the four-wire resistance is subtracted from the two-wire resistance. We have found that  all the contact resistances (including the resistance of the wirebond) are close to $1.18 \pm 0.07~\Omega$. 

\subsection{RF Properties of the Gold Film Electrodes }

\begin{figure}
    \centering
     \includegraphics[width=6.75in]{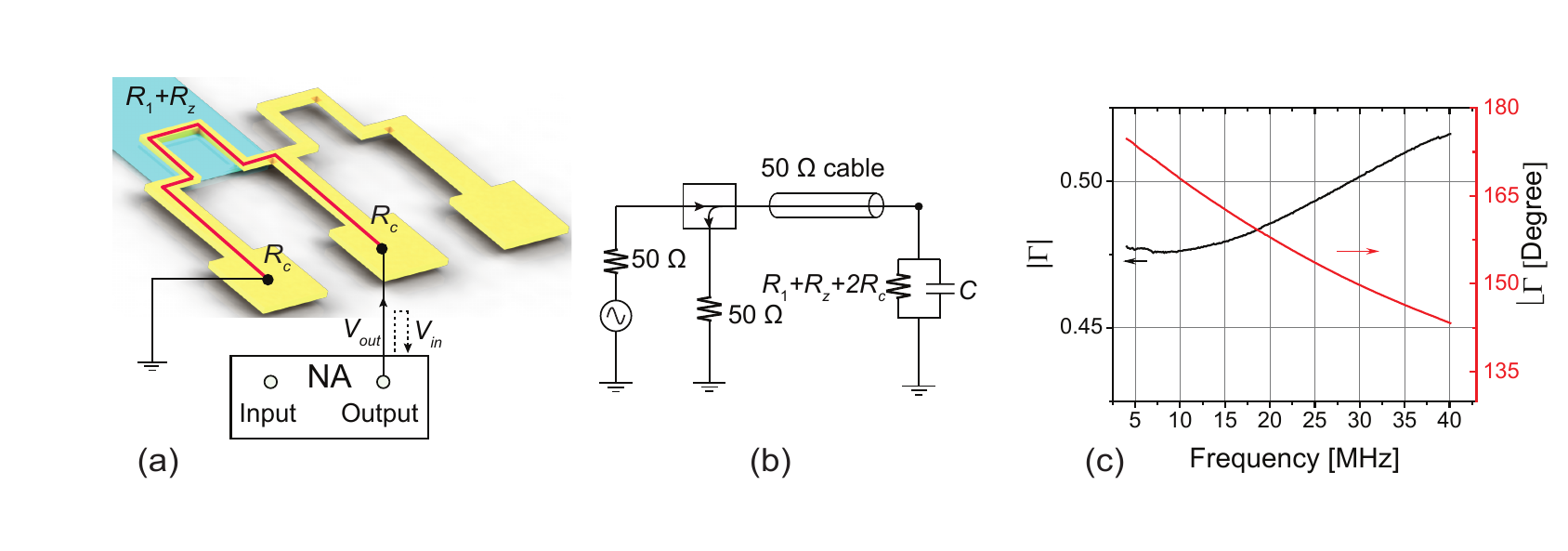}
    \caption{ (a) Schematic diagram of the  setup for measuring the reflection coefficient ($\Gamma$) using a network analyzer (NA).  (b)  Equivalent circuit diagram for (a). (c) The magnitude and phase of  $\Gamma$ as a function of   frequency. We obtain a parasitic capacitance of $C \approx 65~\rm pF$ and resistance of  $R_1+R_{z}+2{R_{c}}  \approx 17.7~ \Omega$ by fitting. } 
    \label{Figure RF}
\end{figure}

We use a network analyzer to measure the RF  reflection coefficient from the electrode. In the reflection measurement, we ground one end of the electrode as shown in the circuit diagram in Fig. \ref{Figure RF}(a). Fig. \ref{Figure RF}(b) shows the equivalent circuit. The reflection coefficient, $\Gamma  =   \frac{Z_L-50}{Z_L+50}$,  is obtained as a function of frequency by sweeping the frequency from $\rm 4~\rm MHz$ to $40~\rm MHz$. Here, $Z_L$ is the load impedance seen by the network analyzer (i.e., $Z_L = \frac{R_1+R_{z}+{2R_{c}}}{1 + j\omega (R_1+R_{z}+{2R_{c}}) C} $ and $Z_0 = 50~ \rm \Omega$). Fig. \ref{Figure RF}(c) shows the magnitude and phase of $\Gamma$.  The parasitic capacitance $C$ and the $R_1+R_{z}+2R_{c}$ are estimated to be $C\approx 65~\rm pF$ and $R_1+R_{z}+2R_{c} \approx 17.7 ~\Omega$, respectively. Therefore, at the highest frequency ($\sim 36$ MHz), the parasitic capacitive impedance to (virtual) ground is $X_C \approx 70 ~\Omega$, indicating that most of the bias current  flows through the nanoresistor in the measurements. We do not correct for the small attenuation at the highest frequency for consistency.

\subsection{Analysis of the Piezoresistance Detection Circuit}

\begin{figure}
    \centering
     \includegraphics[width=6.75in]{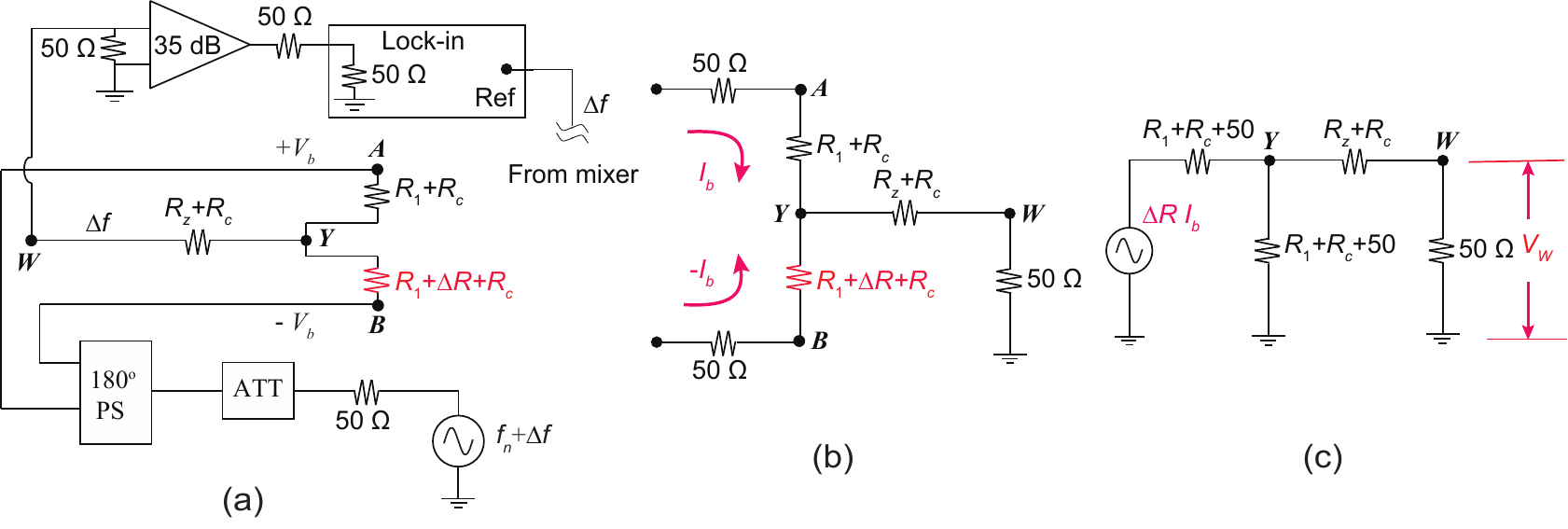}
    \caption{(a) Circuit diagram for electrical detection of piezoresistance.  The bias voltage at frequency $ f_n + \Delta f$ goes   through an attenuator (ATT) and is split into two using a 180$^{\rm o}$ power splitter (PS).  These two voltages marked  $\pm V_b$ are applied to the two outer ports of the bridge, with the corresponding currents going through  the strain gauge and balancing nanoresistor. The piezoresistance $\Delta R$ oscillates at $f_n$, and the piezoresistance signal at the center of the  bridge (at $Y$)  is mixed down to frequency $\Delta f$. The down-converted signal is then ampliﬁed by a pre-ampliﬁer and detected using a lock-in ampliﬁer (SR844, Stanford Research Systems). (b) Simplified  circuit diagram for (a). The $50~\Omega$  resistances on both arms of the bridge represent the source impedances of the signal generator as seen by the device.    (c) Equivalent small signal circuit for detection.  The small current arising from the  piezoresistance  flows into the opposite arm of the bridge and the pre-amplifier.}
    \label{Figure MixCircuit}
\end{figure}

To begin with, we assume that the circuit is balanced when the resonator is off resonance; then, the point $Y$ in the circuit is at virtual ground. Note that, in reality, there is a small degree of imbalance between the two bridge arms due to fabrication, contact resistances, and so on. The circuit is balanced by adjusting the current in each arm  by means of independent variable attenuators. We will thus ignore this imbalance. Under the assumption of a balanced bridge, the current in each arm of the bridge is  ${I_b} (t) \approx  \frac{\sqrt{2}V_b}{(R_1+{R_{c}}+50~ \Omega)}\cos({2 \pi f_n t + 2 \pi \Delta ft})$, both flowing toward point $Y$.  Because of the  mechanical strain at resonance, the transducer  gains a time-dependent resistance change $ \sqrt{2}\Delta R \cos{2 \pi f_n t}$, with $\Delta R \ll R_1$. Given the current in the arm,  a voltage of $ \frac{V_b \Delta R \cos({2\pi\Delta f t})}{(R_1+{R_{c}} +50~ \Omega)}$  develops across $R_1+{R_{c}}+50~\Omega$. To continue the small-signal analysis, we simplify the circuit to that shown in Fig. \ref{Figure MixCircuit}(c) and turn this piezoresistance-based voltage into a voltage source  with the source resistor $R_1+{R_{c}}+50~\Omega$. This voltage then  pushes a current through the parallel resistor combination $(R_{z}+{R_{c}}+50~\Omega)||(R_1+{R_{c}}+50~ \Omega)$.  We find the voltage detected on the $50~\Omega$ input of the preamplifier at point $W$ as
\begin{equation}\label{eq:V_Q}
   {V}_W(t) = \sqrt{2} V_{W} \cos{\left(2\pi\Delta f t\right)} = {\frac{ V_b \Delta R \cos{\left(2\pi\Delta f t\right)} }{2 \left(R_{z} +{R_{c}} + 50~ \Omega\right) \left(R_1+{R_{c}} + 50~ \Omega\right)+\left(R_1+{R_{c}} + 50~ \Omega\right)^2}} \times 50~\Omega.
\end{equation}
After considering the gain $\cal G$ of the preamplifier and the signal division by the $50~\Omega$ resistors, we find the voltage detected  by the lock-in amplifier to be ${ V}_{LI}(t)=\frac{{\cal G}{ V }_W (t)}{2}$.

The same analysis can be performed using the currents flowing into the junction $Y$. We assume as above that, without mechanical motion, the currents flowing into point $Y$ are balanced. When there is mechanical  motion, we use  Kirchhoff's Current Law and express the currents flowing into the junction $Y$ as 
\begin{equation}\label{eq current flowing}
    \frac{{\sqrt{2}} V_b \cos{\left(2 \pi f_n + 2 \pi \Delta f\right)t} - V_Y}{R_1 +R_{c} + 50~ \Omega} = \frac{V_Y}{R_{z} + R_{c}+50~ \Omega} + \frac{V_Y - \Big({-\sqrt{2}}V_b \cos{\left(2 \pi f_n + 2 \pi \Delta f\right)t}\Big)}{R_1 + \sqrt{2}\Delta R \cos{2 \pi f_n t} + R_{c} + 50~ \Omega}.
\end{equation}
The voltage at  junction $V_Y$ can be found as 
\begin{equation}\label{eq V_Y}
     V_Y  \approx \frac{V_b \Delta R \cos{2 \pi \Delta f t} ({R_{z} + R_{c} +50~ \Omega})}{\big[2\left(R_{z} + R_{c} +50~ \Omega\right) + {\left(R_1 + R_{c} +50~ \Omega\right)}\big] \left(R_1 + R_{c} +50~ \Omega\right)}.
\end{equation}
The voltage is then divided by the resistors $R_{z} + R_{c}$ and the 50 $\Omega$ resistor. The end result, namely the voltage $V_W$ detected on the $50~\Omega$ input of the preamplifier, is the same as that in Eq.~(\ref{eq:V_Q}).

Finally, we have repeated a set of measurements using a high impedance pre-amplifier (SR560, Stanford Research Systems) and a slightly lower mix-down frequency of 630 kHz. Since the input impedance of the SR560 (connected to point $W$ in Fig. \ref{Figure MixCircuit}) is $100~\rm M \Omega$, much larger than $50~\Omega$, the  piezoresistance voltage is simply divided into 2 (assuming the circuit is balanced). Indeed, when we compare these measurements with the $50~\Omega$ measurements on the same device under the same conditions, the results match.

\subsection{Electrical Resonance Measurements for Each Eigen-mode}\label{electrical measurements}

\begin{figure}
    \centering
     \includegraphics[width=5.9in]{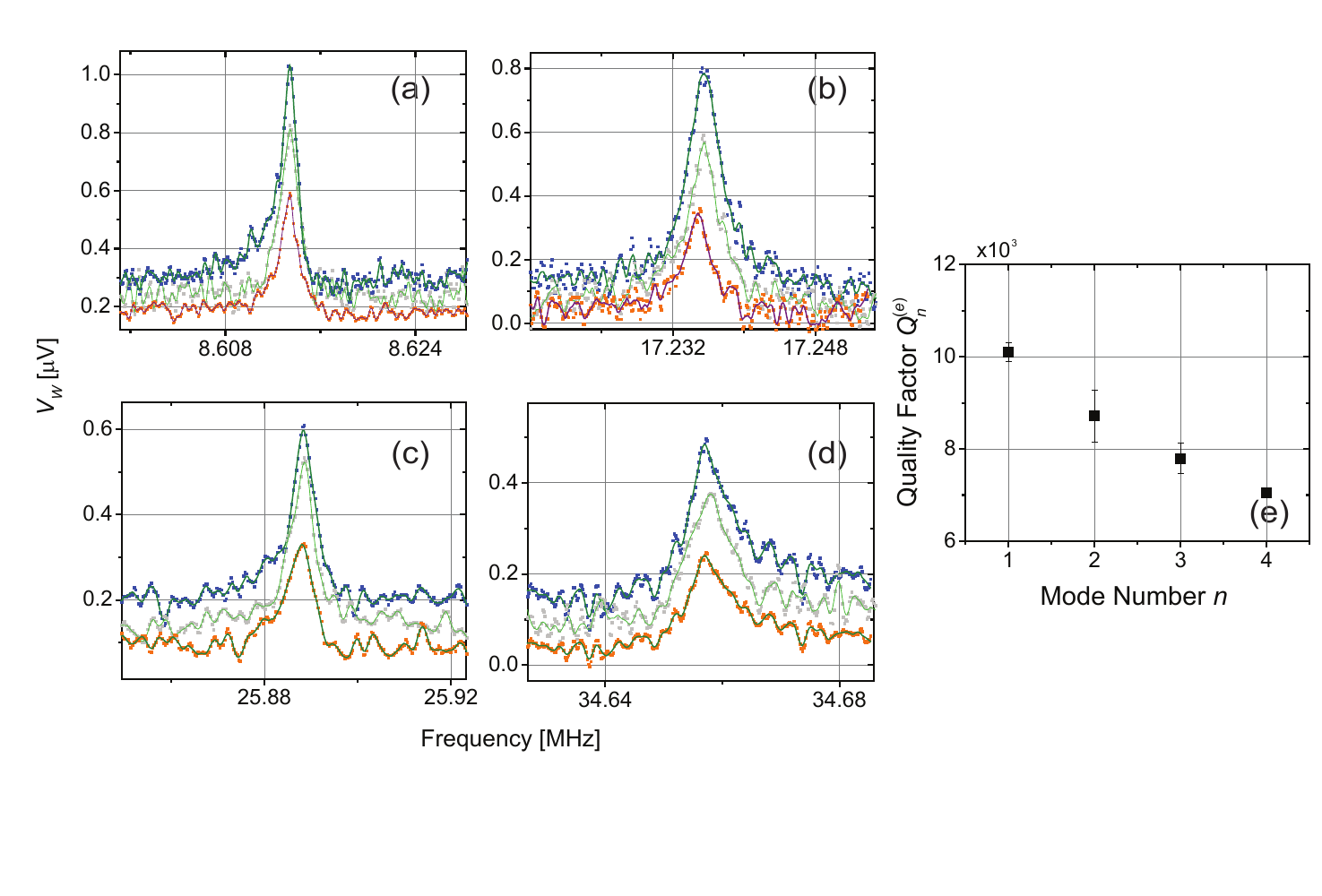}
    \caption{(a-d) Measured rms voltages  at different drives as a function of frequency around the first four eigen-modes of the 30-$\rm \mu$m-long resonator. The drive voltages are 12.3 mV, 11.3 mV, and 8.8 mV for the blue, green, and yellow curves, respectively. (e) Electrical quality factors $Q_n^{(e)}$ as a function of the mode number $n$ for the same resonator.}
    \label{Figure Electr_30}
\end{figure}

\begin{figure}
    \centering
     \includegraphics[width=5.9in]{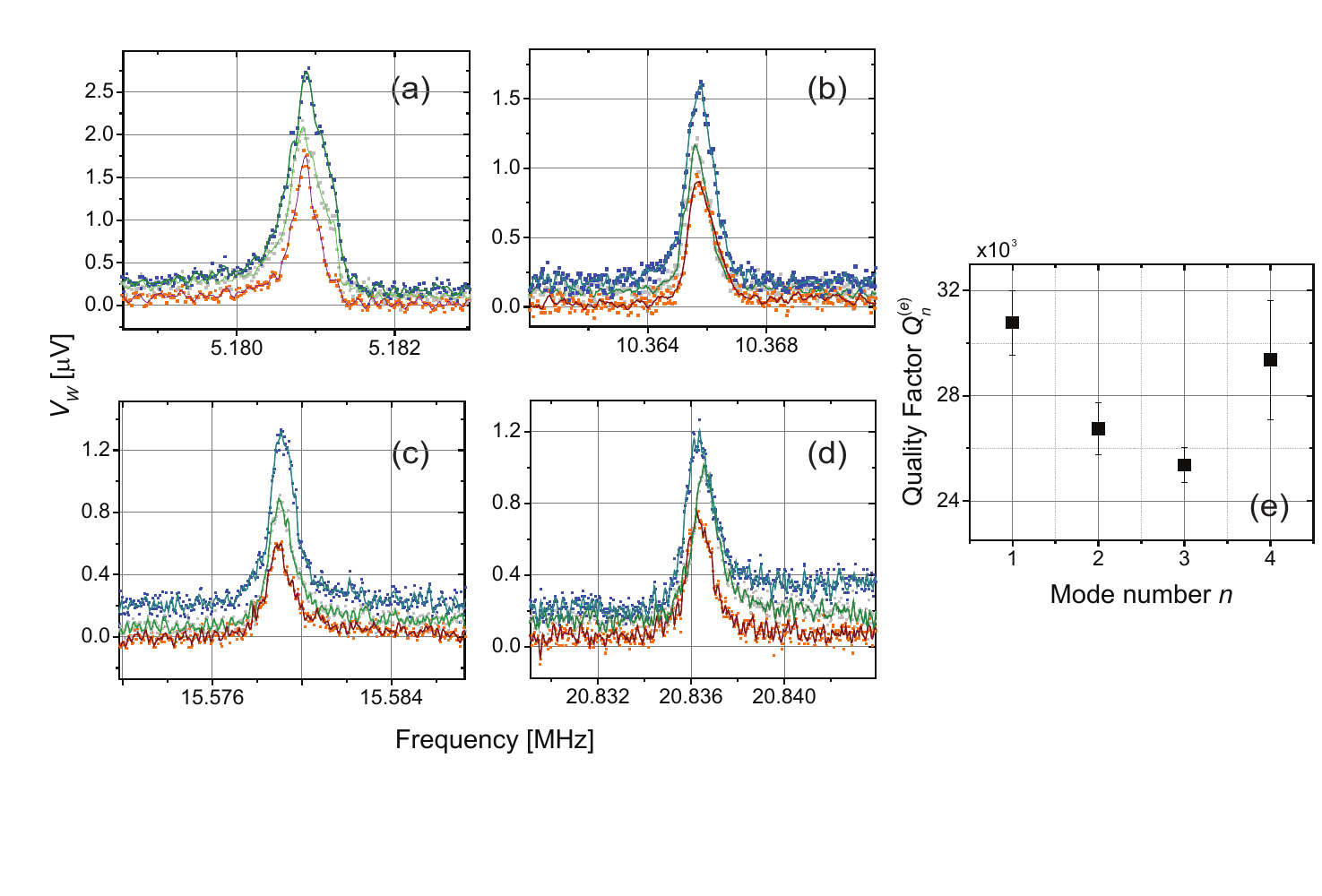}
    \caption{(a-d) Measured rms voltages  at different drives as a function of frequency around the first four eigen-modes of the 50-$\rm \mu$m-long resonator. The drive voltages are 8.7 mV, 7.9 mV, and  7.3 mV for the blue, green, and yellow curves, respectively. (e) Electrical quality factors $Q_n^{(e)}$ as a function of the mode number $n$ for the same resonator. }
    \label{Figure Electr_50}
\end{figure}

\begin{figure}
    \centering
     \includegraphics[width=5.9in]{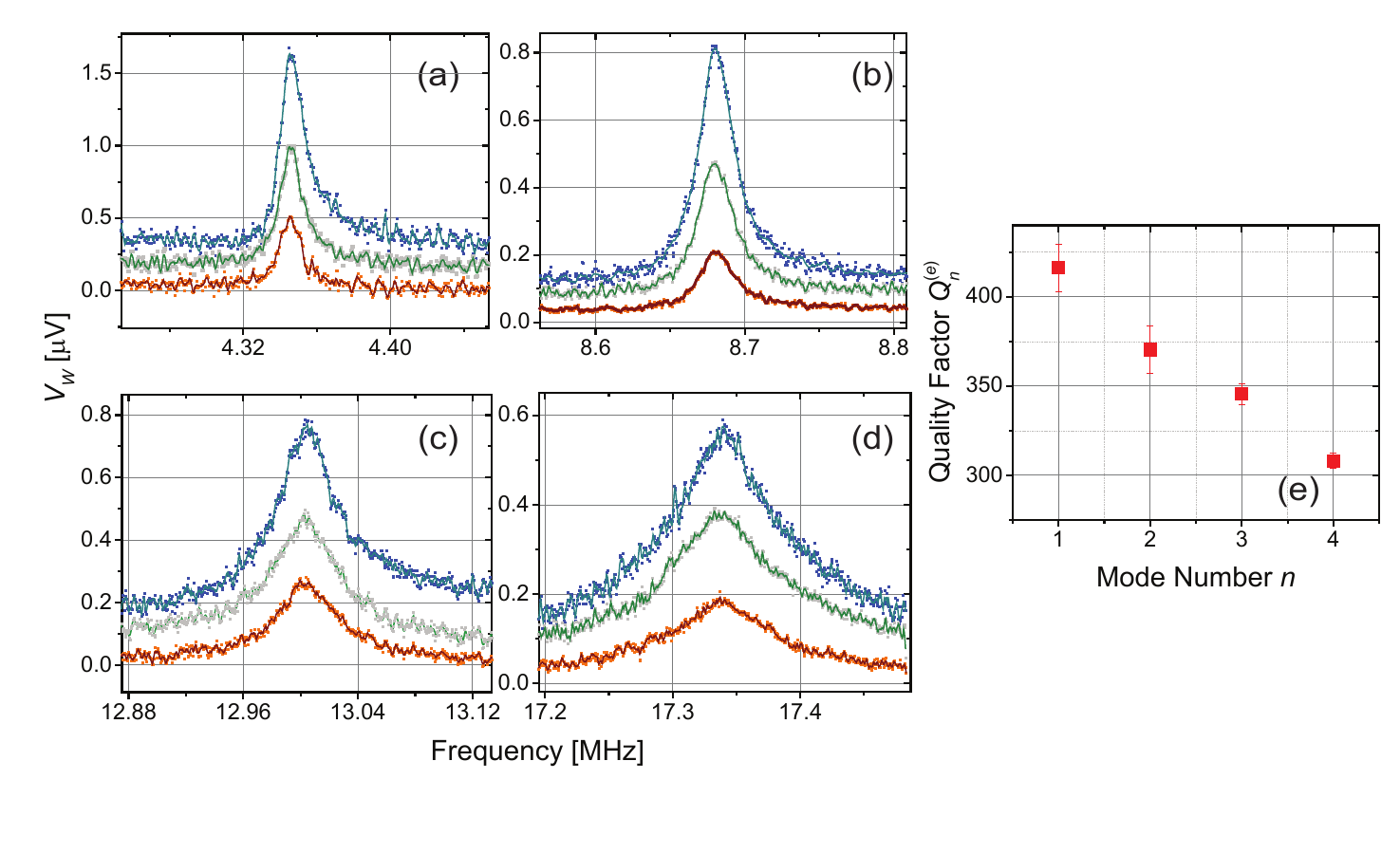}
    \caption{(a-d) Measured rms voltages at different drives as a function of frequency around the first four eigen-modes of the 60-$\rm \mu$m-long resonator. The drive voltages are 60.1 mV, 46.0 mV, and 31.8 mV for the blue, green, and yellow curves, respectively. (e) Electrical quality factors $Q_n^{(e)}$ of different modes for the same resonator.}
    \label{Figure Electr_60}
\end{figure}

Figs. \ref{Figure Electr_30}, \ref{Figure Electr_50}, and \ref{Figure Electr_60} respectively show the detected voltages for  the strain gauges of the 30-$\rm \mu m $-long,  50-$\rm \mu m $-long, and  60-$\rm \mu m $-long resonators at different drives, as the drive frequency is swept through the first four eigen-mode resonances. $V_b$ is kept constant for all eigen-modes of the same resonator. $V_b = 18.5$ mV for the 30-$\rm \mu m $-long resonator, $V_b = 60$ mV for the 50-$\rm \mu m $-long resonator, and $V_b = 150$ mV for the 60-$\rm \mu m $-long resonator.  For each mode, the voltage due to piezoresistance, $ V_{W}$ at $f_n$, is obtained after  background subtraction from the peak amplitude. We note that the baselines at both sides of $f_n$ are not the same, which is taken as a systematic error in the electrical measurement. With  $ V_{W}$ at each amplitude and mode in hand, we determine the $\Delta R$ for different modes of all resonators. 

\begin{figure}
    \centering
     \includegraphics[width=4in]{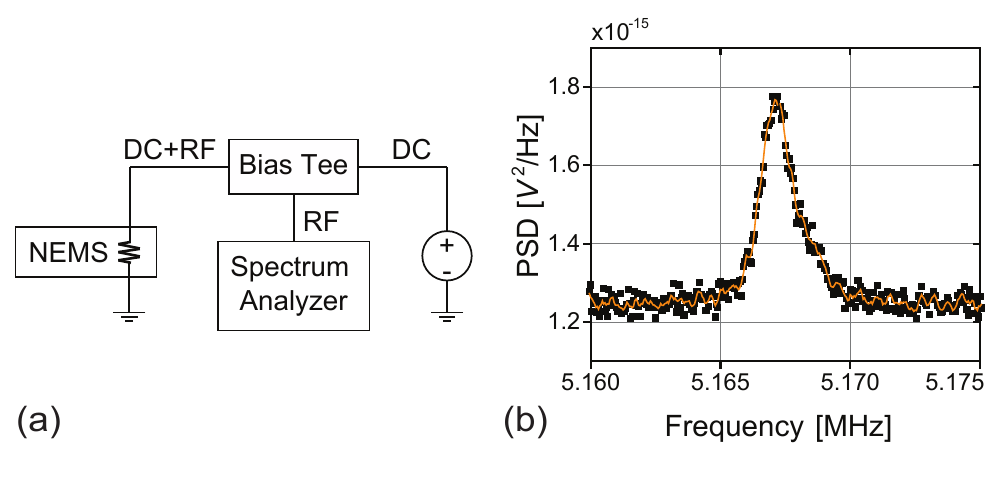}
    \caption{ (a)  The circuit diagram for the thermal noise measurement. The NEMS is biased by a dc voltage of 58.8 mV and its thermal noise is detected by a spectrum analyzer. (b)  PSD of the thermal fluctuations of the 50-$\rm \mu$m-long resonator at its fundamental mode.}
    \label{Figure thermal Elec}
\end{figure}

The thermal fluctuations of the NEMS resonator also provide  sufficient strain to induce detectable piezoresistance on the strain gauge. Fig. \ref{Figure thermal Elec}(a) shows the circuit diagram for the electrical thermal noise measurement. The transducer is biased by a dc voltage of 58.8 $\rm mV$ and the thermal noise from the NEMS device is measured using a spectrum analyzer via a Bias-Tee. Fig. \ref{Figure thermal Elec}(b) displays the noise spectrum of the thermal fluctuations of the 50-$\rm \mu m $-long beam at its fundamental mode frequency detected electrically using  the strain gauge. 

\section{Error Analysis}\label{Error Analysis}

The error bars in Fig. 5 in the main text are determined using standard error analysis. We first estimate the errors along the $x$ and $y$ coordinates for each data point in Figs. 4(a), (b), and (c) in the main text. We then fit  these data sets with errors in both $x$ and $y$ to lines   to obtain  the error bars \cite{york2004unified} for $\gamma$ in each (mode) frequency in Fig. 5. 

To find the $y$ error, we use the equation
\begin{equation}
     \frac{\Delta R}{R_{u}} = \frac{\sqrt{2} V_{W} \left(R_1+R_c + 50~ \Omega\right)^2}{V_b R_t  R_u} \frac{Q^{(o)}}{Q^{(e)}},
    \label{error gamma}
\end{equation}
with all the symbols previously defined.  Here, a source of error is $V_W$, in particular due to the uncertainty in the baseline, as discussed above in Section \ref{electrical measurements}.   The resistances in Eq.~\ref{error gamma}  each have an uncertainty of about $2 - 5\%$, as discussed in the main text. The last sources of error considered are the uncertainties in $Q^{(o)}$ and $Q^{(e)}$.  By propagating all these errors properly, we  determine the $y$ error for each data point in Figs. 4(a), (b) and (c) in the main text, which provides the upper and lower limits of $\frac{\Delta R}{R_{u}}$ at a given $\bar \varepsilon_{xx}$ value. 

The $x$ error for each data point comes from the uncertainty in the $\bar \varepsilon_{xx}$ value. A source of error in $x$  is  the  uncertainty in the undercut length and  the  position of the nanoresistor relative to the suspended region. Both of these factors will result in an error in the strain $\bar \varepsilon_{xx}$ acting on the nanoresistor at a given resonator amplitude.  From  SEM images of several strain gauges from the same batch, we determine these  geometric uncertainties; we then estimate from  FEM simulations the error in $\bar \varepsilon_{xx}$  caused by these geometric uncertainties. We find this error to be about $3\%$ for each data point.  Another source of  $x$ error is the uncertainty in the optical spot position with respect to the anti-node position. When the optical spot is positioned with an error, the amplitude and hence the strain is measured with an error. The  typical error in strain from this source is estimated to be $\lesssim 2\%$, as discussed in Section \ref{Eigen-Mode}. 

\clearpage

\bibliography{Reference.bib}